\documentclass[aps,pre,10pt,showpacs,floatfix,onecolumn,nofootinbib,a4paper]{revtex4-2}
\usepackage{amssymb, amsmath, amsthm, mathtools}
\usepackage{bm}
\usepackage{mathrsfs}
\usepackage{hyperref,bookmark}
\usepackage{graphicx}
\usepackage{epsfig,color}
\usepackage{float}
\usepackage{subcaption}
\usepackage{verbatim}
\usepackage{dcolumn}
\usepackage{tikz}
\usepackage{url}
\usepackage{stmaryrd}

\usepackage[utf8x]{inputenc}
\usepackage[english]{babel}
\usepackage{mathrsfs}
\usepackage{epsfig}
\usepackage{listings}
\usepackage{paralist}

\usepackage{mdwlist}
\usepackage{dsfont}
\usepackage{oldgerm}
\usepackage{geometry}
\usepackage{relsize}
\usepackage{afterpage}
\usepackage{lmodern}
\usepackage{tabularx}
\usepackage{cancel}

\newcommand{\dd}{\mathrm{d}}
\newcommand{\vae}{\varepsilon}
\newcommand{\vt}{\vartheta}
\newcommand{\vp}{\varphi}
\newcommand{\trans}{^\mathsf{T}}
\newcommand{\av}{\bm{a}}
\newcommand{\n}{\bm{n}}
\newcommand{\m}{\bm{m}}
\newcommand{\nper}{\bm{n}_\perp}
\newcommand{\mper}{\bm{m}_\perp}
\newcommand{\cv}{\bm{c}}

\newcommand{\e}{\bm{e}}
\newcommand{\E}{\bm{E}}
\newcommand{\x}{\bm{x}}
\newcommand{\zero}{\bm{0}}
\newcommand{\y}{\bm{y}}

\newcommand{\normal}{\bm{\nu}}
\newcommand{\surface}{\mathscr{S}}
\newcommand{\tube}{\mathscr{T}}
\newcommand{\free}{\mathscr{F}}
\newcommand{\energy}{\mathscr{E}}
\newcommand{\nablas}{\nabla\!_\mathrm{s}}

\newcommand{\ero}{\e_\varrho}
\newcommand{\f}{\bm{f}}
\newcommand{\et}{\e_\vt}
\newcommand{\ep}{\e_\vp}
\newcommand{\bv}{\bm{b}}

\newcommand{\support}{\bm{\gamma}}
\newcommand{\W}{\mathbf{W}}
\newcommand{\F}{\mathbf{F}}
\newcommand{\proj}{\mathbf{P}}
\newcommand{\I}{\mathbf{I}}
\newcommand{\Ln}{\mathbf{L}_{\n}}
\newcommand{\Lm}{\mathbf{L}_{\m}}
\newcommand{\R}{\mathbf{R}}
\newcommand{\C}{\mathbf{C}}

\newcommand{\Q}{\mathbf{Q}}

\newcommand{\tr}{\operatorname{tr}}

\newcommand{\Cf}{\C_{\f}}

\newcommand{\nay}{(\nabla\y)}

\newcommand{\curvature}{(\nablas\normal)}

\newcommand{\body}{\mathscr{B}}
\newcommand{\real}{\mathbb{R}}
\newcommand{\euclid}{\mathscr{E}}
\newcommand{\curve}{\mathscr{C}}
\newcommand{\translation}{\mathscr{V}}
\renewcommand{\frame}{(\e_1,\e_2,\e_3)}
\newcommand{\framem}{(\m,\mper,\e_3)}
\newcommand{\framec}{(\e_x,\e_y,\e_z)}
\newcommand{\framen}{(\n,\nper,\normal)}
\newcommand{\framero}{(\ero,\et,\ep)}
\newcommand{\slab}{\mathsf{S}}

\newcommand{\tangent}{\bm{t}}
\newcommand{\tangentper}{\bm{t}_\perp}
\newcommand{\pnormal}{\bm{n^\ast}}
\newcommand{\binormal}{\bm{b}}
\newcommand{\jump}[1]{{\llbracket{#1}}\rrbracket}

\newcommand{\vc}[1]{\bm{#1}}

\renewcommand{\geq}{\geqslant}

\begin{document}
	\title{Ridge energy for thin  nematic polymer networks}
	\author{Andrea Pedrini}
	\email{andrea.pedrini@unipv.it}
	\author{Epifanio G. Virga}
	\email{eg.virga@unipv.it}
	\affiliation{Dipartimento di Matematica, Universit\`a di Pavia, Via Ferrata 5, 27100 Pavia, Italy}

	\date{\today}

\begin{abstract}
Minimizing the elastic free energy of a thin sheet of nematic polymer network among smooth isometric immersions is the strategy purported by the mainstream theory. In this paper, we broaden the class of admissible spontaneous deformations: we consider \emph{ridged} isometric immersions, which can cause a sharp ridge in the immersed surfaces. We propose a model to compute the extra energy distributed along such ridges. This energy comes from bending; it is shown under what circumstances it scales quadratically with the sheet's thickness, falling just in between stretching and bending energies. We put our theory to the test by studying the spontaneous deformation of a disk on which a radial hedgehog was imprinted at the time of crosslinking. We predict the number of folds that develop in terms of the degree of order induced in the material by external agents (such as heat and illumination).
\end{abstract}
\maketitle

\section{Introduction}\label{sec:intro}
Nematic elastomers are rubber materials with a fluid-like component constituted by elongated, rod-like molecules appended to the crosslinked polymer strands that form the background matrix. The fluid component is ordered as nematic liquid crystals can be, which makes these solid materials very susceptible to external stimuli, such as heat, light, and environmental humidity. The prompt response to these stimuli, so characteristic of liquid crystals, once transferred to the solid matrix, makes it possible to do work and change the shape of bodies with no direct contact. The possible technological applications of these materials are boundless (see, for example, the papers \cite{harris:large,kowalski:pixelated,kowalski:voxel,babakhanova:liquid,zeng:light,brannum:deformation,van_oosten:glassy,van_oosten:bending,van_oosten:printed}, and above all the review \cite{white:programmable}), but a number of theoretical challenges remain open \cite{warner:topographic}; this paper is concerned with one of them.

The order established in the material by the mutual interaction of nematic molecules is described by a scalar order parameter, representing the degree of molecular alignment, and a director, representing the average direction of alignment. Actually, there are two sets of these order parameters, namely, the pair $(s_0,\m)$ for the reference configuration of the rubber matrix, which here will be taken to be the configuration where the crosslinking takes place, and the pair $(s,\n)$ for the current (deformed) configuration, the one the rubber matrix takes on in response to an applied stimulus (more details are given in Sect.~\ref{sec:energies} below). The director $\n$ can be tied to the deformation of the body in several ways, the spectrum going from complete independence to complete enslaving. Following the terminology introduced in \cite{white:programmable}, we call \emph{nematic polymer networks} the nematic elastomers in which the crosslinking in so tight that the nematic director remains enslaved to the deformation;\footnote{This name has not yet met with universal acceptance. Some also say that these are liquid crystal \emph{glasses} \cite{he:making,he:programmed,modes:disclination,plucinsky:programming}, while others prefer to say that they are simply nematic elastomers with a \emph{locked} (or \emph{frozen}) director \cite{cirak:computational}.} these are the specific nematic elastomers treated here. The reason for this choice will soon become clear.

The most striking manifestation of the ability of nematic polymer networks to perform changes in shape is perhaps achieved when they are thin sheets. We represent one such sheet as a slab $\slab$ of thickness $2h$ extending  in the reference configuration on both sides of a flat surface $S$. The director $\m$ is \emph{blueprinted} on $S$ (in its own plane), uniformly reproduced across the thickness, with a given scalar order parameter $s_0$. External stimuli may act on the degree of order, changing $s_0$ into $s$, in a programmable way. The system is thus carried out of equilibrium and a deformation ensues, for the free energy to attain a minimum under the changed circumstances.

An elastic free-energy density, $f_e$, is available for bulk materials in three space dimensions since the pioneering work \cite{blandon:deformation} (a comprehensive introduction to the subject is offered by the landmark book \cite{warner:liquid}); it is delivered by the ``trace formula'', derived from assuming an anisotropic Gaussian distribution for the polymer chains that constitute the rubber matrix.\footnote{Critiques have been moved to this formula. A noticeable improvement was achieved in \cite{kutter:tube} through a successful extension of Edward's \emph{tube} model \cite{edwards:theory} for entangled rubber elasticity. Here, however, we shall abstain from dwelling any further on possible extensions of the trace formula, as desirable as these may be.} This formula features both the deformation $\f$ of the three-dimensional body $\body$ occupied by the material and measures of anisotropy in both reference and current configurations of $\body$ (see Sect.~\ref{sec:energies}.) For a sufficiently thin slab $\slab$, however, one's desire is to reduce $f_e$ to a function of the mapping $\y$ that only changes the flat reference mid surface $S$ into a curved surface $\surface$  in the current configuration.

In a nematic polymer network, for which $f_e$ eventually depends only on $\f$, such a dimension reduction was performed in \cite{ozenda:blend} by revisiting (and extending) a standard method of the theory of plates, known as the Kirchhoff-Love hypothesis \cite{ozenda:kirchhoff}. As expected, this method delivers a surface elastic energy with two components, a \emph{stretching} energy $f_s$ scaling like $h$, and a bending energy $f_b$ scaling like $h^3$; $f_s$ depends only on the two-dimensional stretching (or metric) tensor $\C:=(\nabla\y)\trans(\nabla\y)$, while $f_b$ also depends on the invariant measures of curvature of $\surface$ and the relative orientation of $\n$ in the frame of principal directions of curvature. Not only do $f_s$ and $f_b$ scale differently with $h$, they are also basically different things. By Gauss' \emph{theorema egregium} \cite[p.\,139]{stoker:differential}, the Gaussian curvature $K$ of $\surface$ is fully determined by the metric tensor $\C$, thus deserving the name of \emph{intrinsic} curvature. As a consequence, $f_s$ depends only on the intrinsic curvature, whereas $f_b$ also depends on extrinsic measures of curvature, relating on how $\surface$ is embedded in three-dimensional space. Studying the complete equilibrium problem, where $f_b$ is treated on the same footing as $f_s$ has proven so far difficult. A number of strategies have been devised to circumvent the energy coupling. 

For moderately curved surfaces $\surface$ and sufficiently thin slabs $\slab$, for which $f_b$ can be neglected relative to $f_s$, the energy minimizing shapes are \emph{isometric immersions}\footnote{Here we may be guilty of some abuse of language, as the metric induced on $\surface$ by $\y$ differs from the Euclidean metric on $S$ whenever $\C\neq\I$. However, we may think of endowing $S$ with the metric described by any given symmetric, positive tensor $\C$ and ask whether $S$, so endowed, can be immersed in three-dimensional Euclidean space preserving the metric. In this sense, which will always be understood here, the word \emph{isometry} is justified.} of the metric tensor $\C_0$ that minimizes $f_s$. The search for such immersions corresponding to a variety of imprinted $\m$ fields has been the subject of a vast, elegant literature (see, among others, \cite{modes:gaussian,modes:negative,plucinsky:programming,mostajeran:curvature,mostajeran:encoding,mostajeran:frame,kowalski:curvature,warner:nematic}) This may seem to solve the direct \emph{morphic mechanics} problem for nematic polymer networks, namely, how to identify the shapes produced by a certain imprinted director field $\m$. More difficult (and less visited), but affordable is the \emph{inverse} problem of assigning $\m$ so as to produce a desired shape upon stimulation \cite{griniasty:curved}.

As reassuring as this picture may appear, things are unfortunately more complicated than they look like: there are at least two conflicting, as it were, unresolved issues. A \emph{smooth} isometric immersion with prescribed metric tensor $\C_0$ may altogether fail to exist in the large. On the other hand, if we renounce the smoothness requirement for the immersion, the number of admissible solutions may easily become too large.\footnote{In Sect.~\ref{sec:hedgehog} below, we shall provide plenty of examples for continuous isometric immersions with continuous $\m$, but discontinuous $\normal$.}

A remedy for the first issue was proposed by the theory of \emph{geometric elasticity} \cite{aharoni:geometry,aharoni:universal}. If the target metric corresponding to $\C_0$ is geometrically incompatible with a smooth immersion, this theory proposes to replace it with the one that minimizes an appropriate $L^2$-distance from it. It is a viable approximation, if you do not wish to renounce regularity.

A remedy for the second issue would be provided by a selection criterion that single out one shape out of many, preferably on energetic grounds. Here the essential question is:  what extra energy should be attached to a singular shape? This is the avenue taken here. We allow $\surface$ to have \emph{ridges}, that is, lines along which the outer unit normal $\normal$ suffers a discontinuity. As for the extra energy cost to be associated with a ridge, we extract it from the bending energy density $f_b$. We conceive a ridge as a limiting tight fold, for which  we justify an expression for a \emph{ridge} linear density $f_r$, which depends (in a symmetric way) on the traces $\normal_1$ and $\normal_2$ of the unit normal $\normal$ on both sides of the ridge. It turns out that in our theory $f_r$ scales like $h^2$, just in between $f_s$ and $f_b$, so that $f_r$ becomes the effective substitute for $f_b$. This justifies an approximation alternative to  geometric elasticity: finding piecewise isometric $C^2$-immersions that minimize the total ridge energy.

The \emph{vicarious} nature of our theory is to be stressed from the start. The real (still unresolved) challenge is minimizing the total elastic energy of a thin nematic polymer network, with both stretching and bending contributions. Failing to do so, we find it expedient to \emph{replace} a distributed bending energy with a concentrated one, which is simpler than the former (and scales differently too), but is not \emph{ad hoc}. Were the same replacement adopted for the Euler \emph{elastica}, one would find a similarly viable theory.

The paper is organized as follows. In Sect.~\ref{sec:energies}, we recall both stretching and bending energies for nematic polymer networks, as they emerged from the dimension reduction of the bulk energy density delivered by the trace formula. Section~\ref{sec:elastica} plays the role of an intermezzo in our development: there we show how to destructure the classical \emph{elastica} in a chain of rigid rods connected by articulated joints (edges) encapsulating the bending energy of the parent body. We shall see how this simplified model is capable of capturing the known qualitative behavior of the \emph{elastica}, thus paving the way to our ridge construction.  In Sect.~\ref{sec:ridge}, we construct the ridge energy as a limit of the bending energy entrapped in a folded sheet. Section~\ref{sec:ridges_embeddings} is concerned with the general equations that govern piecewise $C^2$-immersions with discontinuities of the unit normal field $\normal$ concentrated along smooth curves; such \emph{ridged} isometric immersions are the shape competing for a minimum in our theory. In Sect.~\ref{sec:hedgehog}, we consider some special symmetric ridged isometric immersions that mimic the folds generated  in a disk when the imprinted director field $\m$ is the radial hedgehog; we compute the ridge energy that acts as an obstruction to the proliferation of folds and we determine their optimal number. Section~\ref{sec:conclusions} is where we draw our conclusions and comment on other possible uses of our theory. The paper is closed by an appendix, where we illustrate a geometric construction apt to produce the analytic solution proposed in Sect.~\ref{sec:hedgehog} for the ridged isometric immersion of a hedgehog.

\section{Stretching and bending energies}\label{sec:energies}
In this section, we recall the outcomes of the dimension reduction method applied in \cite{ozenda:blend} to the trace formula of the \emph{neo-classical}
theory for nematic elastomers (for which we refer the reader to Chap.~6 of \cite{warner:liquid}). Two director fields feature in this theory; these are $\m$, defined in the reference configuration $\body$ of the body, and $\n$, defined in the current configuration $\f(\body)$ obtained from $\body$ through the deformation $\f$. $\body$ is a region in three-dimensional Euclidean space $\euclid$ and $\f:\body\to\euclid$ is a diffeomorphism of $\body$.

The directors $\m$ and $\n$ represent the average alignment of the elongated molecules appended to the rubber polymeric matrix in the reference and current configurations. They are properly defined through the tensorial measures of anisotropy that  characterize the end-to-end Gaussian distribution of polymer strands. These are the polymer \emph{step tensors} $\Lm$ and $\Ln$, in the reference and current configurations, respectively, which, following \cite{verwey:elastic} and \cite{nguyen:theory}, we write as 
 \begin{subequations}\label{eq:L_definitions}
	\begin{equation}\label{eq:L_m}
	\Lm:=\mathfrak{a}_0(\I+s_0\m\otimes\m)
	\end{equation}
	and 
	\begin{equation}\label{eq:L_n}
	\Ln:=\mathfrak{a}(\I+s\n\otimes\n).
	\end{equation}
\end{subequations}
Here $\I$ is the identity (in three-dimension space), $\mathfrak{a}_0$ and $\mathfrak{a}$ are fixed positive parameters (representing the persistence lengths perpendicular to $\m$ and $\n$, respectively), $s_0$ and $s$ are nematic scalar order parameters, which can be expressed as $s_0=r_0-1$ and $s=r-1$ in terms of the ratios $r_0$ and $r$ of the parallel (along $\m$ and $\n$) and perpendicular (across $\m$ and $\n$) step chain lengths in the reference and current configurations, respectively.

The neo-classical theory of nematic elastomers expresses the elastic free-energy density $f_e$ (per unit volume in the reference configuration) as
\begin{equation}\label{eq:energy_density}
f_e:=\frac12k\tr(\F\trans\Ln^{-1}\F\Lm),
\end{equation}
where $\F:=\nabla\f$ is the deformation gradient and $k>0$ is an elastic modulus (which scales linearly with both absolute temperature and number density of polymer chains). This is usually called the \emph{trace formula}.

In nematic elastomers, $\n$ and $\F$ are fully independent. In contrast, in nematic polymer networks, $\n$ is enslaved to $\F$. In these materials, with which we are concerned in this paper, the director field $\m$ is \emph{blueprinted} in the elastic matrix \cite{modes:blueprinting} and conveyed by the deformation into $\n$, which is thus delivered by
\begin{equation}\label{eq:n}
\n=\frac{\F\m}{|\F\m|}.
\end{equation}
In general, elastomers are \emph{incompressible}, and so $\F$ must satisfy
\begin{equation}\label{eq:incompressibility_F}
\det\F=1.
\end{equation}
Both \eqref{eq:n} and \eqref{eq:incompressibility_F} will be enforced as constraints on all admissible deformations $\f$ of $\body$.

With $\m$ (and $s_0$) imprinted in the reference configuration at the time of crosslinking and $\n$ enslaved to the deformation, the only residual freedom lies with $s$, which can be changed by either thermal or optical stimuli. For example, by heating the sample above the crosslinking temperature, we reduce the nematic order of the chains, so that $s<s_0$; this in turns induces a spontaneous deformation so as to minimize the total elastic free energy. Thus, $s$ can be regarded as the \emph{activation parameter} of our theory, driven by external stimuli. For definiteness, we shall assume that both $s_0$ and $s$ range in the interval $(-1,1)$.

It was shown in \cite{ozenda:blend} that by use of \eqref{eq:L_definitions} and \eqref{eq:n} $f_e$ can be given the following form
\begin{equation}\label{eq:F_definition}
f_e=\frac12k\frac{\mathfrak{a}_0}{\mathfrak{a}}F(\Cf),
\end{equation}  
where $\Cf:=\F\trans\F$ is the right Cauchy-Green tensor associated with the deformation $\f$ and 
\begin{equation}\label{eq:bulk_energy_density}
F(\Cf)=\tr\Cf+\frac{s_0}{s+1}\m\cdot\Cf\m-\frac{s}{s+1}\frac{\m\cdot\Cf^2\m}{\m\cdot\Cf\m}.
\end{equation}
The properties of this function will illuminate the role of $s$ as activation parameter.

As a consequence of \eqref{eq:incompressibility_F}, $\Cf$ is also subject to the constraint 
\begin{equation}
\label{eq:det_C_f=1}
\det\Cf=1.
\end{equation}
The tensors $\Cf$ that make $F(\Cf)$ stationary subject to \eqref{eq:det_C_f=1} are solutions to the equation
\begin{equation}
\label{eq:F_equilibrium_equation}
\frac{\partial F}{\partial\Cf}=\lambda\frac{\partial}{\partial\Cf}(\det\Cf),
\end{equation} 
where $\lambda$ is a Lagrange multiplier. It is not difficult to see that this equation reduces to 
\begin{equation}
\label{eq:F_equilibrium_equation_reduced}
\I+\frac{1}{s+1}\left(s_0+s\frac{\m\cdot\Cf^2\m}{(\m\cdot\Cf\m)^2}\right)\m\otimes\m-\frac{s}{s+1}\frac{1}{\m\cdot\Cf\m}(\Cf\m\otimes\m+\m\otimes\Cf\m)=\lambda\Cf^{-1}.
\end{equation}
It follows from \eqref{eq:F_equilibrium_equation_reduced} that changing $\m$ into $\Q\m$, for any orthogonal tensor $\Q$, transforms a solution $\Cf$ into $\Q\Cf\Q\trans$, which makes any solution $\Cf$ of \eqref{eq:F_equilibrium_equation_reduced} an isotropic tensor-symmetric-valued function of $\m$. By the representation theorem of such functions \cite{wang:new_II} and \eqref{eq:det_C_f=1}, we know that $\Cf$ must have the form
\begin{equation}
\label{eq:Cf_representation}
\Cf=\lambda_f^2\m\otimes\m+\frac{1}{\lambda_f}(\I-\m\otimes\m),
\end{equation} 
for some $\lambda_f\in\real^+$.
Making use of \eqref{eq:Cf_representation} in \eqref{eq:F_equilibrium_equation_reduced}, we readily conclude that 
\begin{equation}
\label{eq:lambda_f}
\lambda=\frac{1}{\lambda_f}\quad\text{and}\quad\lambda_f=\sqrt[3]{\frac{s+1}{s_0+1}}.
\end{equation} 
By expressing $F(\Cf)$ in terms of $\lambda_f$ with the aid of \eqref{eq:Cf_representation}, it is easy to show that for $\lambda_f$ as in \eqref{eq:lambda_f} this function attains its unique minimum.

Thus, when $s<s_0$, the spontaneous deformation induced in the material would be a contraction along $\m$, accompanied by a dilation in the plane orthogonal to $\m$, to preserve the volume.\footnote{Clearly, still according to \eqref{eq:lambda_f}, for $s>s_0$, which is achieved upon cooling the sample below the crosslinking temperature, the material would expand along $\m$ and contract transversely.}  Of course, it remains to be seen whether, for an assigned $\m$, a deformation with a metric that minimizes $F$ locally is indeed geometrically compatible in the large; differently put, whether there is an isometric immersion in three space dimensions of the desired target metric $\Cf$ as in \eqref{eq:Cf_representation}.

\begin{figure}[h]
	\begin{center}
		\includegraphics[width=.5\linewidth]{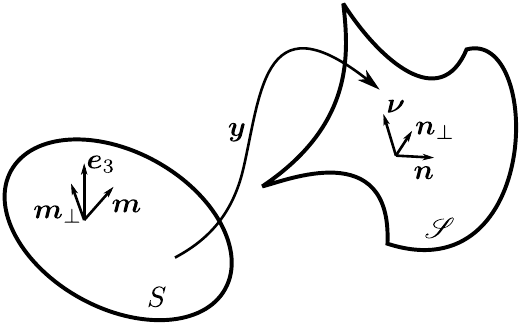}
	\end{center}
	\caption{\label{fig:sketch} A flat surface $S$ in the $(x_1,x_2)$ plane of a fixed Cartesian frame $\frame$ is deformed by the mapping $\y$ into a smooth surface $\surface$ embedded in three-dimensional Euclidean space $\euclid$. The blueprinted orientation is denoted by $\m$ in the reference configuration and by $\n$ in the current one; $\e_3$ is the outer unit normal to $S$, while $\normal$ is the outer unit normal to $\surface$;  correspondingly, $\mper:=\e_3\times\m$ and $\nper:=\normal\times\n$.}
\end{figure}

Here we are interested in thin sheets and in the appropriate dimension reduction of $F(\Cf)$ to be attributed to the mid surface $S$ of the slab $\slab$ of thickness $2h$. Formally, $S$ is a flat region in the $(x_1,x_2)$ plane of a fixed Cartesian frame $\frame$ and $\slab$ is the set in three-space defined as $\slab:=\{(\x,x_3)\in S\times[-h,h] \}$. The mapping $\y:S\to\euclid$ describes the deformation of $S$ into the surface $\surface=\y(S)$ in the deformed slab $\f(\slab)$; we shall assume that $\y$ is of class $C^2$ and that $\m$ is a two-dimensional field imprinted on $S$, so that $\m\cdot\e_3\equiv0$ (see Fig.~\ref{fig:sketch}).\footnote{In $\slab$, $\m$ is extended uniformly away from $S$, so as to be independent of the $x_3$ coordinate.}

The (two-dimensional) deformation gradient has the following general representation,
\begin{equation}\label{eq:nabla_y_representation}
\nabla\y=\av\otimes\m+\bv\otimes\mper,
\end{equation}
where $\mper:=\e_3\times\m$. In \eqref{eq:nabla_y_representation}, $\av$ and $\bv$ are vector fields defined on $S$; they live in $\translation$, the translation space of $\euclid$, and are everywhere tangent to $\surface$. It follows from \eqref{eq:nabla_y_representation} that the two-dimensional stretching tensor $\C$ is represented as
\begin{equation}\label{eq:C}
\C=(\nabla\y\trans)(\nabla\y)=a^2\m\otimes\m+\av\cdot\bv(\m\otimes\mper+\mper\otimes\m)+b^2\mper\otimes\mper,
\end{equation}
where $a^2:=\av\cdot\av$ and $b^2:=\bv\cdot\bv$. We shall require that $S$ is \emph{inextensible}, which amounts to the constraint $|\av\times\bv|=1$. Thus, since $\det\C=a^2b^2-(\av\cdot\bv)^2=|\av\times\bv|^2$, we shall require that 
\begin{equation}\label{eq:det_C=1}
\det\C=1.
\end{equation}
Under this constraint, the outer unit normal $\normal$ to $\surface$ will be delivered by
\begin{equation}\label{eq:normal}
\normal=\av\times\bv.
\end{equation}
Applying \eqref{eq:n} to the present setting, we obtain that
\begin{equation}
\label{eq:n_y_setting}
\n=\frac{(\nabla\y)\m}{|(\nabla\y)\m|},
\end{equation}
and so we may write $\av=a\n$ and define $\nper:=\normal\times\n$, so that the frame $\framen$ is oriented as $\framem$ (see Fig.~\ref{fig:sketch}).

In \cite{ozenda:blend}, we extended the classical Kirchhoff-Love hypothesis \cite{ozenda:kirchhoff} to obtain a dimension reduction of $F(\Cf)$ in \eqref{eq:bulk_energy_density}, that is, a method that convert $f_e$ in \eqref{eq:F_definition} into a surface energy-density (to be integrated over $S$). As standard in the theory of plates, such a surface energy is delivered by a polynomial in odd powers of $h$, conventionally truncated so as to retain the first two relevant ones, the first and the third power. The former is the \emph{stretching} energy $f_s$, accounting for the work done to alter distances and angles in $S$, while the latter is the \emph{bending} energy $f_b$, accounting for the work done to fold $S$. Thus, dropping the scaling constant $\frac12k\frac{\mathfrak{a}_0}{\mathfrak{a}}$, which has the physical dimensions of an energy per unit volume, we can write
\begin{equation}
\label{eq:energy_splitting}
f_e=f_s+f_b+O(h^5),
\end{equation}
where (to within an inessential additive constant)
\begin{subequations}
\begin{eqnarray}
f_s&=&\frac{2h}{s+1}\left(\tr\C+s_0\m\cdot\C\m+\frac{s}{\m\cdot\C\m}\right),\label{eq:f_s}\\
f_b&=&\frac{2h^3}{3}\left\{2(8H^2-K)+\frac{1}{s+1}\left[\left(\frac{3s}{a^2}-a^2s_0-\tr\C\right)K-\frac{4s}{a^2}(2H-\kappa_n)\kappa_n\right] \right\}.\label{eq:f_b}
\end{eqnarray}
\end{subequations}
Here $H$ and $K$ are the mean and Gaussian curvatures of $\surface$, defined as
\begin{equation}
\label{eq:H_K_definitions}
H:=\frac12\tr\curvature\quad\text{and}\quad K:=\det\curvature
\end{equation}
in terms of the (two-dimensional) curvature tensor $\nablas\normal$, and
\begin{equation}
\label{eq:11}
\kappa_n:=\n\cdot\curvature\n.
\end{equation}
The (scaled) total elastic free energy then reduces to the functional
\begin{equation}
\label{eq:free_energy_functional}
\free[\y]:=\int_S(f_s+f_b)\dd A,
\end{equation}
where $A$ is the area measure. 

A perturbation approach to the minimization of $\free$ is justified when the length scale associated with the average radius of curvature of $\surface$ is large compared to $h$, which is the smallest length in the system: then $f_s$ and $f_b$ are well scale-separated and the latter can be viewed as a higher-order correction to the former. In this approach, it is justified to ask what stretching tensor $\C_0$ would minimize $f_s$, the leading term in $\free$. The answer is easily obtained \cite{ozenda:blend}, that is,
\begin{equation}\label{eq:C_stationary}
\C_0=\lambda_1^2\m\otimes\m+\lambda_2^2\mper\otimes\mper,
\end{equation}
where
\begin{equation}\label{eq:lambda_1_2}
\lambda_1:=\sqrt[4]{\frac{s+1}{s_0+1}}\quad\text{and}\quad\lambda_2=\frac1\lambda_1.
\end{equation}

A deformation $\y$ for which \eqref{eq:C_stationary} is valid is an \emph{isometric immersion}; it minimizes the (leading) stretching energy. The problem is then whether such immersions do exist and how many they are.\footnote{Here we do not discuss boundary conditions, which may have a disquieting role. We think only of shapes in space, which are then defined to within arbitrary translations and rotations.} This is when the bending energy comes to play. If there are no isometric immersions, it means that $f_s$ must be \emph{blended} with $f_b$ and more elaborate minimizing shapes $\surface$ must be sought for, presumably exhibiting regions where the average radius of curvature is not much larger than $h$. On the other hand, if there are many isometric immersions, we may hope to use the bending energy as a selection criterion, choosing the isometric immersion with the least bending energy.

Both scenarios, however, are overoptimistic. The first, because minimizing the blended energy is not an easy task, also numerically, as the functional $\free$ depends on the second as well as the first gradient of $\y$. The second, because the \emph{two-step} minimization, which unleashes $f_b$ over the minimizers of $f_s$, may actually turn out to be rather disappointing; for example, only spheres are allowed among surfaces $\surface$ with positive $K$, if one insists in minimizing $f_b$ uniformly \cite{ozenda:blend}.

So far we have thought of isometric immersions as smooth mappings. The regularity issue now becomes relevant and tackling it opens up new perspectives. As shown in Sect.~\ref{sec:ridges_embeddings}, one can easily incur in  many an isometric immersion  by relaxing the requirement that $\y$ be $C^2$. We shall consider mappings $\y$ that are piecewise $C^1$, with $\nabla\y$ allowed to jump across one or several \emph{ridges}, which will be assumed to be  smooth curves of class $C^1$. One such mapping is a piecewise isometric immersion if $(\nabla\y)\trans(\nabla\y)\equiv\C_0$ on the \emph{whole} domain  $S$, despite the discontinuities of $\nabla\y$ across ridges.

The issue with such \emph{ridged} immersions is that they effectively encapsulate a bending energy in the ridges across which the outer unit normal to $\surface$ jumps abruptly. In Sec.~\ref{sec:ridge}, by regarding each of these ridges as a tight \emph{fold} with continuous principal curvatures extending over a fixed length $\vae\sim h$,  we shall extract out of $f_b$ an elastic  ridge energy-density (per unit length) $f_r$. This energy, which scales like $h^2$, will replace $f_b$ in the purpose of mitigating the multiplicity of ridged isometric immersions. It will form the basis of our (simplified) model for nematic polymer networks. Before all this, to motivate better our moderately unritual approach to isometric immersions, we pause briefly and apply a similar approach to a well established elastic problem, that of the \emph{elastica}.\footnote{We owe to the critical remarks of a reviewer the addition of the following section.}

\section{Intermezzo: Disembodied \emph{elastica}}\label{sec:elastica}
As is well known, the \emph{elastica} is a one-dimensional continuum body represented by an inextensible curve in space endowed with bending stiffness. We may think of this as the mid line of a thin three-dimensional body, extended in one direction much more than in the other two. In its simplest incarnation, the elastic energy stored in the \emph{elastica} is
\begin{equation}
	\label{eq:elastica_energy}
	\energy_b=\frac12B\int_0^L\kappa^2\dd\xi,
\end{equation}
where $B>0$ is the bending modulus, $L$ is the length of the mid line, $\xi$ the arc-length parameter and $\kappa$ the curvature.

\begin{figure}[h]
	\centering
		\includegraphics[width=.4\textwidth]{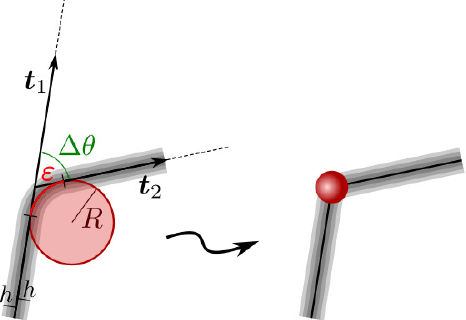}
		\caption{Sketchy justification of the reduction of $\energy_b$ to $\energy_e$. The fibers of an \emph{elastica} are bent over a fixed length $\vae$ along a circle of radius $R$. We extract from them the relevant portion of bending energy $\energy_b$ and attribute it to an edge where the unit tangent $\tangent$ suffers an abrupt disalignment, from $\tangent_1$ to $\tangent_2$, measured by the angle $\Delta\theta=\vae/R$.}
		\label{fig:elastica_fibers}
\end{figure}

Now, instead of distributing the bending energy all along the mid line, we concentrate it in a finite number of places, $N$. In each of these, as shown in Fig.~\ref{fig:elastica_fibers}, we imagine to explore the three-dimensional body at the length scale of the diameter $2h$ of its (circular) cross-section. The mid line  
is bent by an angle $\Delta\theta$ along a circular arc of fixed length $\vae$, bearing an elementary \emph{edge} energy
\begin{equation}
	\label{eq:elastica_edge}
	\Delta E:=\frac12\frac{B}{\vae}(\Delta\theta)^2,
\end{equation}
which easily follows from \eqref{eq:elastica_energy}. Next, we replace the \emph{elastica} by a chain of $N$ rigid rods, each of length $L/N$, connected to the adjoining ones by an \emph{edge}, to which we assign the energy $\Delta E$ in \eqref{eq:elastica_edge}, with $\vae$ fixed and $\Delta\theta$ expressing the mismatch of the concurring rods (see Fig.~\ref{fig:elastica_fibers}). In a way, here we go backwards along Hencky's route \cite{hencky:uber}. We extract from the energy of Euler's \emph{elastica} the energy that Hencky's model attributes to an articulated system of rigid rods connected through torsional springs.\footnote{Here we are \emph{not} interested in taking $N$ large to see how Hencky's discrete model approaches Euler's continuum model. For this, we refer the reader to a rich, still flourishing literature, of which the following papers represent just a few remarkable examples, \cite{alibert:extensional,alibert:convergence,bruckstein:epi-convergence,espanol:euler,scholtes:variational}.} Admittedly, this is a cruder model, which is however amenable to a simple numerical study revealing the main qualitative features of classical solutions, as we now proceed to show by example.

\begin{figure}
	\centering
		\begin{subfigure}[t]{.1\textwidth}
		\centering
		\includegraphics[width=.37\textwidth]{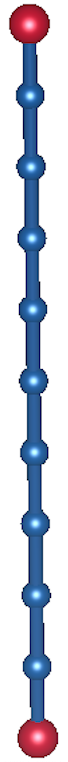}
		\caption{$\alpha=1$}
	\end{subfigure}
	\begin{subfigure}[t]{.11\textwidth}
		\centering
		\includegraphics[width=.85\linewidth]{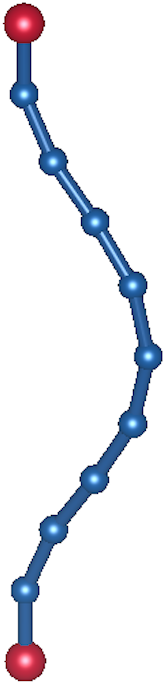}
		\caption{$\alpha=0.9$}
	\end{subfigure}\quad
	\begin{subfigure}[t]{.11\textwidth}
	\centering
	\includegraphics[width=1.05\linewidth]{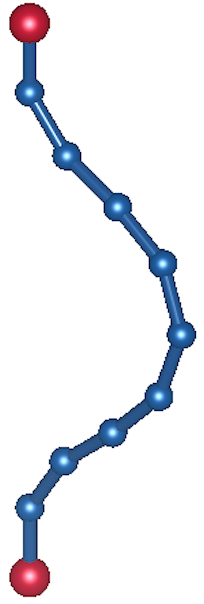}
	\caption{$\alpha=0.8$}
	\end{subfigure}\quad
	\begin{subfigure}[t]{.11\textwidth}
	\centering
	\includegraphics[width=1.05\linewidth]{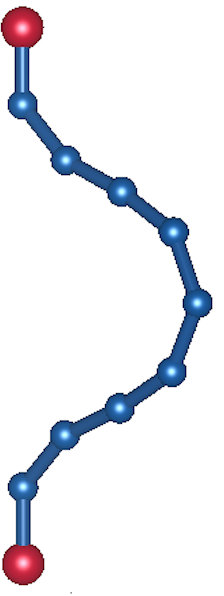}
	\caption{$\alpha=0.7$}
\end{subfigure}\quad
\begin{subfigure}[t]{.11\textwidth}
	\centering
	\includegraphics[width=.85\linewidth]{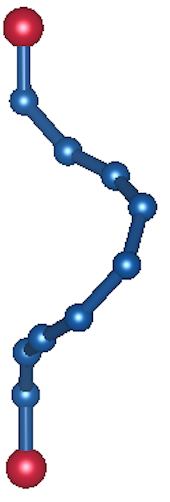}
	\caption{$\alpha=0.6$}
\end{subfigure}\quad
\begin{subfigure}[t]{.11\textwidth}
	\centering
	\includegraphics[width=.95\linewidth]{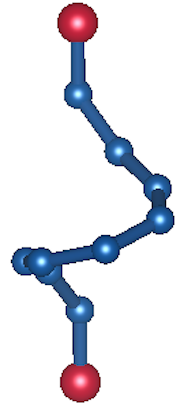}
	\caption{$\alpha=0.5$}
\end{subfigure}\quad
\begin{subfigure}[t]{.11\textwidth}
	\centering
	\includegraphics[width=.7\linewidth]{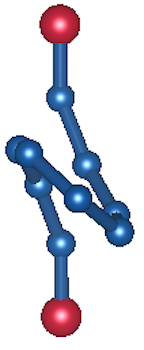}
	\caption{$\alpha=0.4$}
\end{subfigure}
	\begin{subfigure}[t]{.11\textwidth}
		\centering
		\includegraphics[width=1.2\linewidth]{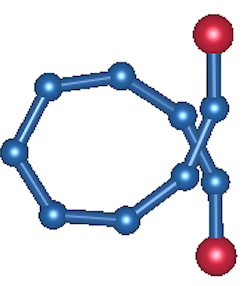}
		\caption{$\alpha=0.3$}
\end{subfigure}
	\caption{A gallery of absolute energy minimizers for $\energy_e$ in \eqref{eq:elastica_chain_energy} subject to the constraints \eqref{eq:elastica_constraints} are computed for $N=10$ and decreasing values of the (normalized) end-to-end separation $\alpha$ of a disembodied clamped \emph{elastica}. Red balls represent the clamped ends, while blue balls represent the articulated edges where two adjacent rods concur. The buckling transition to a twisted (out-of-plane) configuration can be placed in the interval $0.6<\alpha<0.7$. The system is seen from different perspectives in different panels to help the reader visualizing the twisted configurations. The Python code (with adjustable parameters $\alpha$ and $N$) that produced these stable equilibrium configurations can be found (and run) in \cite{pedrini:elastica}. All picture were drawn using  VESTA software~\cite{momma:vesta}.}
	\label{fig:elastica_gallery}
\end{figure}

First, we write the total energy of the chain as
\begin{equation}
	\label{eq:elastica_chain_energy}
	\energy_e:=\frac12\frac{B}{h}\sum_{i=1}^{N-1}\arccos^2(\tangent_{i+1}\cdot\tangent_i),
\end{equation}
where we have set $\vae=h$, which represents the smallest length scale in the model. Here $\tangent_i$ is the unit tangent vector along the $i$-th rod in the chain. In particular, we want to study the equilibrium problem of a chain whose first and last rods are clamped one on top of the other at a distance $a<L$ (see Fig.~\ref{fig:elastica_gallery}), so that, in a Cartesian frame $\framec$, $\tangent_1=\tangent_N=\e_z$.
Letting $\tangent_i$ be represented in spherical coordinates as
\begin{equation}
	\label{eq:elastica_t_i}
	\tangent_i=\sin\theta_i\cos\phi_i\e_x+\sin\theta_i\sin\phi_i\e_y+\cos\theta_i\e_z
\end{equation}
with $\theta_i\in[0,\pi]$ and $\phi_i\in[0,2\pi)$, we easily give $\energy_e$ in \eqref{eq:elastica_chain_energy} the following form
\begin{equation}
	\label{eq:elatica_chain_energy_rewritten}
	\energy_e=\frac12\frac{B}{h}\sum_{i=1}^{N-1}\arccos^2\left(\sin\theta_{i+1}\sin\theta_i\cos(\phi_{i+1}-\phi_i)+\cos\theta_{i+1}\cos\theta_i\right),
\end{equation}
subject to the constraints
\begin{equation}
	\label{eq:elastica_constraints}
	\frac{1}{N}\sum_{i=1}^N\cos\theta_i=\frac{a}{L}=:\alpha,\quad \sum_{i=1}^N\sin\theta_i\cos\phi_i=0,\quad \sum_{i=1}^N\sin\theta_i\sin\phi_i=0,
\end{equation}
which (with $\theta_1=\theta_N=0$) ensure that the boundary conditions are met.

This chain of articulated rods, which pays an elastic disalignment cost at the edges, is our \emph{disembodied elastica}. We minimized numerically $\energy_e$ in \eqref{eq:elatica_chain_energy_rewritten} subject to \eqref{eq:elastica_constraints} for decreasing values of $0<\alpha<1$; we used a stochastic gradient descent method for which a Python code is available in \cite{pedrini:elastica} (and can be run with parameters $\alpha$ and $N$ of the user's choice).  The outcomes of our computations are shown in Fig.~\ref{fig:elastica_gallery} for $N=10$. Upon decreasing $\alpha$, we first find the  absolute minimum of $\energy_e$ in a planar configuration (one of the infinitely many possible was selected with a bias for $\phi=0$), but as soon as $\alpha$ becomes smaller than a critical value $\alpha_\mathrm{c}$ the absolute minimum of $\energy_e$ is attained on either of two (equally energetic) out-of-plane configurations exhibiting a spontaneous twist, which persists and grows upon further reducing $\alpha$. Such a buckling instability, which injects chirality into the system, was already found in \cite{miyazaki:analytical} by a bifurcation analysis of the equilibrium equations for the parent \emph{elastica} (illustrated, in particular, in their Figs.~3 and 4); in our parameterization, their bifurcation point lies at $\alpha_\mathrm{c}\doteq0.63$ (corresponding to the case where the torsional stiffness vanishes, as is implicit in \eqref{eq:elastica_energy}). Our estimate ($0.6<\alpha_\mathrm{c}<0.7$) is clearly approximate, but the qualitative agreement between disembodied  and full fleshed \emph{elasticae} is undeniable.

Reassured by this agreement, achieved with just a \emph{small} number of rods, in the following section, we shall propose a similar simplified representation for the bending energy of nematic polymer networks.\footnote{Again, we stress that here we are not interested in establishing convergence of the discrete model to the continuum model. Our aim is extracting judiciously from the latter the energy fit for the former. Only a few rods may suffice to establish a subtle qualitative feature, such as the twisting instability of a clamped \emph{elastica}.} What here are edges, will there be ridges. 

\section{Ridge energy}\label{sec:ridge}
\begin{figure}[h]
		\centering
		\includegraphics[width=.5\textwidth]{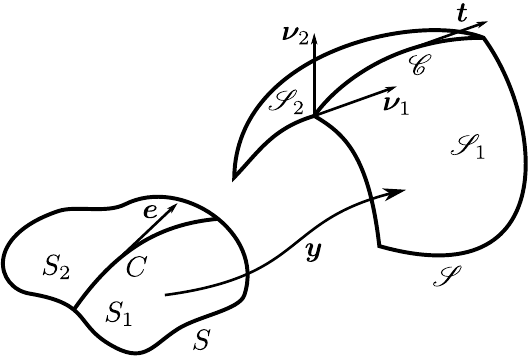}
		\caption{The reference surface $S$ is split by a smooth curve in two sides, $S_1$ and $S_2$, which a deformation $\y$, continuous through $C$, but with discontinuous gradient, maps into the sides $\surface_1$ and $\surface_2$ of the ridge $\curve$. The unit tangent vector $\tangent$ to $\curve$ is related by \eqref{eq:t_from_e} to the unit tangent vector $\e$ to $C$.}
		\label{fig:Fig2_a}
	\end{figure}

In this section, we describe how we envision what at large scale is a ridge on $\surface$: this is generated by what at short scale  is a sharp bend of $\slab$;   we shall  derive from $f_b$ in \eqref{eq:f_b} the energy that can be associated with it. We start from the large-scale perspective. Let $C$ be a smooth (plane) curve on $S$ (say, of class $C^1$) splitting $S$ in two sides, $S_1$ and $S_2$ and let $\e$ be a unit tangent vector to $C$ (see Fig.~\ref{fig:Fig2_a}).
A deformation $\y:S\to\euclid$, continuous across $C$ but with discontinuous gradient $\nabla\y$, must obey the following kinematic compatibility condition,
\begin{equation}
\label{eq:y_jump_condition}
\jump{\nabla\y}\e=\zero,
\end{equation}
where the jump $\jump{(\cdot)}:=(\cdot)_2-(\cdot)_1$ is taken on the two sides $S_2$  and $S_1$ of $C$. The deformation $\y$ transforms $C$ into a smooth curve $\curve$ on $\surface$ along which the outer unit normal $\normal$ is discontinuous; we call $\curve$ a \emph{ridge} of $\surface$. We shall denote by $\normal_1$ and $\normal_2$ the traces of $\normal$ taken on the sides $\surface_1$ and $\surface_2$, connecting smoothly (in a $C^1$-fashion) along $\curve$ and corresponding to the sides $S_1$ and $S_2$ of $C$, respectively.\footnote{Of course, one could easily envision more complicated splittings of $S$ (and, correspondingly, more complicated ridges on $\surface$). Here, we prefer  to keep things simple at first, and then generalize in an assumption the result suggested by our simple construction.} 

We designate by $\support$ the parameterization of $\curve$ in the arc-length $\ell$ and correspondingly we call $\tangent(\ell)=\support'(\ell)$ its unit tangent vector; $\tangent$ is related to $\e$ through the equation
\begin{equation}
\label{eq:t_from_e}
\tangent=\frac{\nay_i\e}{|\nay_i\e|},
\end{equation}
where by \eqref{eq:y_jump_condition} $i$ can take either value $i=1,2$, according to the side upon which $\curve$ is approached.

The mapping $\y$ is a \emph{ridged} isometric immersion if
\begin{equation}
\label{eq:ridged_embedding_definition}
\jump{\nay\trans\nay}=\zero,
\end{equation}
meaning that the stretching tensor $\C$ is continuous across $C$.\footnote{Possibly chosen as in \eqref{eq:C_stationary}, though this is not necessary for the validity of our conclusions here.} The field $\m$ is taken to be continuous across $C$, but by \eqref{eq:n_y_setting} $\n$ generally fails to be continuous across $\curve$. However, because of the identities
\begin{equation}
\label{eq:jump_identities}
\jump{\nay\m\cdot\nay\m}=0,\quad\jump{\nay\e\cdot\nay\e}=0,\quad\text{and}\quad\jump{\nay\e\cdot\nay\m}=0,
\end{equation}
which follow from \eqref{eq:ridged_embedding_definition} and the continuity of $\m$, we arrive  at
\begin{equation}
\label{eq:jump_n_dot_t}
\jump{\n\cdot\tangent}=0,
\end{equation}
so that the projection of $\n$ along the ridge must be the same on both its sides.

Now, we turn to the short-scale perspective. Much in tune with the geometric construction employed in disembodying the \emph{elastica} in the preceding section, we imagine that a ridge $\curve$ results from a sharp bend in the  mid surface $\surface$ of $\f(\slab)$ extending over a fixed length $\vae$ comparable with the shortest length scale $h$ in the model. To establish a clear connection between such a short-scale  bent surface and the large-scale ridge $\curve$, we need digress slightly.

\begin{figure}[h]
	\centering
	\begin{subfigure}[t]{.6\textwidth}
		\centering
		\includegraphics[width=\textwidth]{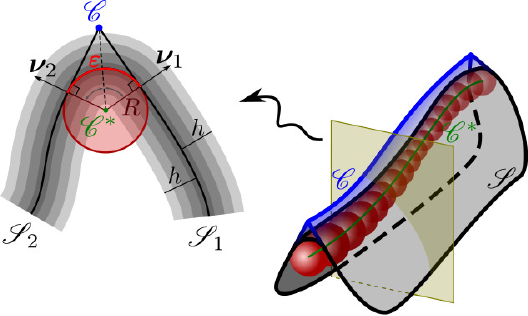}
		\caption{An illustration of the tube construction adopted here to justify equation \eqref{eq:ridge_energy_density} for the linear density $f_r$ of ridge energy. $R$ is the radius of the sphere that generates the tube surface $\tube$ as its center runs along the curve $\curve^\ast$, $\vae$ is the width of $\tube$ over which the mid surface $\surface$ is bent. The ridge $\curve$ is constructed as explained in the text, see especially equation \eqref{eq:C_representation}.}
		\label{fig:tube_generation}
	\end{subfigure}
	\begin{subfigure}[t]{.4\textwidth}
		\includegraphics[width=\textwidth]{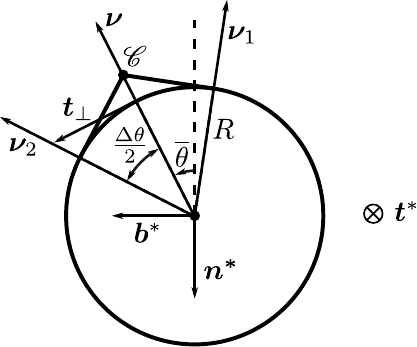}
		\caption{Cross-section of the tube surface $\tube$ described by \eqref{eq:tube_description}. The outer unit normal $\normal$ to the tube is expressed through \eqref{eq:tube_normal} in terms of the principal normal $\pnormal$ and the binormal $\binormal^\ast$ of $\curve^\ast$. The normals $\normal_1$ and $\normal_2$ delimit the tube connecting $\surface_1$ and $\surface_2$. The unit vector $\tangent^\ast$ entering the page is tangent to the curve $\curve^\ast$ described in space by the center of the spheres that generate $\tube$.}
		\label{fig:Fig2_b}
	\end{subfigure}
	\caption{The tube construction for a single ridge.}
	\label{fig:tube_construction}
\end{figure}

Consider  a \emph{tube} surface $\tube$ (see, for example, pp.\,649--650 of \cite{gray:modern}) generated by the motion of a sphere of (possibly variable) radius $R$ whose center travels along a curve $\curve^\ast$, see Fig.~\ref{fig:tube_generation}.
Such a surface can be parameterized as follows,
\begin{equation}
\label{eq:tube_description}
\bm{p}(\xi,\theta)=\support^\ast(\xi)+R(-\cos\theta\pnormal(\xi)+\sin\theta\binormal^\ast(\xi))\quad\text{for}\quad0\leqq\xi\leqq L^\ast\quad\text{and}\quad0\leqq\theta\leqq2\pi,
\end{equation}
where $\support^\ast$ is the parameterization of $\curve^\ast$ in the arc-length $\xi$, and $(\tangent^\ast,\pnormal,\binormal^\ast)$ designates its Frenet-Serret frame.

The situation we envision is shown  in Fig.~\ref{fig:Fig2_b}. The tube surface $\tube$ connects two  components, $\surface_1$ and $\surface_2$, of the mid surface $\surface$  over a length $\vae$, which is related to the (finite) angle $\Delta\theta=|\theta_2-\theta_1|$ delimited by the connecting normals, $\normal_1$ and $\normal_2$, through 
\begin{equation}\label{eq:epsilon}
	\vae=R\Delta\theta.
\end{equation}	
At this scale, we identify the curve $\curve$ that will serve as ridge at a coarser scale by taking  the intersection on the $(\pnormal,\binormal^\ast)$ plane between the lines orthogonal to $\normal_1$ and $\normal_2$, as shown in Fig.~\ref{fig:Fig2_b}. In this construction, the radius $R$ of the traveling sphere, as well as the angles $\theta_1$ and $\theta_2$ can be taken as (smooth) functions of the arc-length parameter $\xi$; for the derivatives of these functions, we shall assume that both
\begin{equation}
	\label{eq:smallness_assumption}
\vae|\theta_1'|\ll1\quad\text{and}\quad\vae|\theta_2'|\ll1.
\end{equation}
In the parameters $(\xi,\theta)$, the outer unit normal $\normal$ to $\tube$ reads as
\begin{equation}
\label{eq:tube_normal}
\normal=-\cos\theta\pnormal(\xi)+\sin\theta\binormal^\ast(\xi),
\end{equation}
and we can represent $\curve$ as
(see Fig.~\ref{fig:Fig2_b}) 
\begin{equation}
	\label{eq:C_representation}
	\support=\support^\ast+\frac{R}{\cos\left(\frac{\Delta\theta}{2}\right)}(-\cos\overline{\theta}\pnormal+\sin\overline{\theta}\binormal^\ast),
\end{equation}
where we have set $\overline{\theta}:=\frac12(\theta_1+\theta_2)$. It is now a simple matter to show that, if in addition to \eqref{eq:smallness_assumption} we also assume that the curvature $\kappa^\ast$ and torsion $\tau^\ast$ of $\curve^\ast$ are such that
\begin{equation}
	\label{eq:smallness_assumption_bis}
	\vae\kappa^\ast\ll1\quad\text{and}\quad\vae|\tau^\ast|\ll1,
\end{equation}
the curves $\curve$ and $\curve^\ast$ are nearly parallel and the curvature $\kappa$ of $\curve$ can  be identified with the curvature $\kappa^\ast$ of $\curve^\ast$: they differ by terms vanishing with $\vae$ as do the arc-length parameters $\ell$ and $\xi$.\footnote{It may also be  worth noting that, by \eqref{eq:epsilon}, inequalities \eqref{eq:smallness_assumption} imply that  $|R'\Delta\theta|=R|\Delta\theta'|=\vae\Delta\theta'/\Delta\theta\ll1$. Our estimates are based on the assumption that $\Delta\theta$ stays finite. Should it be infinitesimal instead, our reasoning would still be valid, provided that $\Delta\theta$ dominates the infinitesimals in both \eqref{eq:smallness_assumption} and \eqref{eq:smallness_assumption_bis}.} With this identification, 
it is not difficult to see that the curvature tensor of $\tube$ is given by (see also \cite[p.\,650]{gray:modern})
\begin{equation}
\label{eq:tube_curvature_tensor}
\nablas\normal=\frac{\kappa\cos\theta}{1+R\kappa\cos\theta}\tangent\otimes\tangent+\frac{1}{R}\tangentper\otimes\tangentper,
\end{equation}
where $\tangentper:\normal\times\tangent$, so that 
\begin{equation}
\label{eq:tube_H_K}
H=\frac12\left(\frac1R+\frac{\kappa\cos\theta}{1+R\kappa\cos\theta}\right)\quad\text{and}\quad K=\frac1R\frac{\kappa\cos\theta}{1+R\kappa\cos\theta}.
\end{equation}
The area element is correspondingly delivered by
\begin{equation}
\label{eq:tube_area_element}
\dd A=(1+R\kappa\cos\theta)\dd\ell\dd\sigma,
\end{equation}
where $\dd\sigma=R\dd\theta$.

The aim of this construction is to extract from $f_b$ the bending energy concentrated in a jump of $\normal$ that is to be assigned  to $\curve$ as an energy distributed over its length. To this end, we recall that,  at the leading order in $\vae\kappa$, $f_b$ in \eqref{eq:f_b} can be estimated from \eqref{eq:tube_curvature_tensor} as
\begin{subequations}
	\label{eq:tube_h_H_and_K}
\begin{gather}
	\label{eq:tube_H_and_K}
H=\frac{\Delta\theta}{2\vae} +O(\vae\kappa),\quad K=\frac{\Delta\theta}{\vae}\kappa\cos\theta+O(\vae\kappa),\quad\text{so that}\quad\frac{K}{H^2}=O(\vae\kappa),\\ \text{and}\quad(2H-\kappa_n)\kappa_n=\left(\frac{\Delta\theta}{\vae}\right)^2(\n\cdot\tangent)^2(\n\cdot\tangentper)^2+O(\vae\kappa).
\end{gather}
\end{subequations}
Making use of \eqref{eq:tube_h_H_and_K} in \eqref{eq:f_b}, we arrive at 
\begin{equation}
\label{eq:tube_f_b}
f_b=\frac83\frac{h^3}{\vae^2}\left(1-\frac{s}{s+1}\frac{1}{a^2}(\n\cdot\tangent)^2(\n\cdot\tangentper)^2\right)(\Delta\theta)^2+O(\vae\kappa),
\end{equation}
where we recall that $a^2=\m\cdot\C\m$. Integrating this in the tube delimited by $\theta_1$ and $\theta_2$, since by \eqref{eq:jump_n_dot_t} and the constraint $\n\cdot\n=1$ both $(\n\cdot\tangent)^2$ and $(\n\cdot\tangentper)^2$ are continuous across $\curve$ and can be taken as independent of $\sigma$ over the tube, by \eqref{eq:tube_area_element} we estimate a single ridge energy $\free_r$ as
\begin{equation}
\label{eq:ridge_energy}
\free_r:=\int_0^L\int_0^\vae f_b\dd\ell\dd\sigma =\frac83h^2\int_0^L\left(1-\frac{s}{s+1}\frac{1}{a^2}(\n\cdot\tangent)^2(\n\cdot\tangentper)^2\right)(\Delta\theta)^2\dd\ell+O(h^2\kappa ),
\end{equation}
where, on physical grounds, we have taken $\vae=h$, for $h$ is the smallest length scale meaningful in our model. This justifies reducing 
$\free_r$ to a line integral along $\curve$ with density (per unit length)
\begin{equation}
\label{eq:ridge_energy_density}
f_r:=\frac83h^2\arccos^2(\normal_1\cdot\normal_2)\left(1-\frac{s}{s+1}\frac{1}{a^2}(\n\cdot\tangent)^2(\n\cdot\tangentper)^2\right),
\end{equation}
which is the main outcome of our tube construction.

If, at a  length scale larger than $h$, $\surface_1$ and $\surface_2$ are isometrically immersed   and meet at the ridge $\curve$, this latter is endowed with the extra energy $\free_r$ in \eqref{eq:ridge_energy}. In case of multiple ridges $\curve_j$, we shall simply \emph{assume} that 
\begin{equation}
\label{eq:ridge_energy_multiple}
\free_r:=\sum_{j=1}^N\int_{\curve_j}f_r\dd\ell,
\end{equation}
where $N$ is the total number of ridges present on $\surface$. This is the total \emph{ridge energy} that we shall assign here to a ridged isometric immersion.

Two comments are in order. First, by direct inspection of \eqref{eq:ridge_energy_density}, it is evident that for $s>0$ the ridge energy-density would promote an alignment of $\n$ at $\frac\pi4$ with the ridge  (on both adjoining sides), whereas for $s<0$ it would equally promote an alignment either parallel or orthogonal to the ridge. Second, and more importantly, since $\free_r$ scales like $h^2$, it dominates over the bending energy distributed over the smooth components $\surface_k$ of $\surface$ isometrically immersed in three-space. Thus, in our theory $\free_r$ becomes the effective substitute for the bending energy.
\footnote{This scaling follows from the choice of taking the width $\vae$ of $\tube$ comparable with $h$ in our tube construction of a ridge. Although this seems the most natural choice, it remains questionable. Different energy concentration mechanisms could indeed suggest different scalings laws for the ridge energy (such as the power $h^{8/3}$ contemplated in \cite{lobkovsky:scaling}). Be it as it may, for our purposes we only need make sure that  $\free_r$ scales with a power of $h$ that dominates (for small $h$) over $h^3$, which is the scaling of the bending energy.}

In the following section, we shall write the equations that describe a ridged immersion in a special representation. An example of $\free_r$ will be computed explicitly in Sect.~\ref{sec:hedgehog}.

\section{Representing ridged isometric immersions}\label{sec:ridges_embeddings}
Away from possible point defects (the only ones allowed here), the director field $\m$ imprinted on $S$ and its orthogonal companion $\mper$ are assumed to have continuous gradients, which can be represented as 
\begin{equation}\label{eq:connector}
\nabla\m=\mper\otimes\cv\quad\text{and}\quad\nabla\mper=-\m\otimes\cv
\end{equation}
in terms of the planar \emph{connector} field $\cv$ \cite{ozenda:blend}. We shall use the frame $\framem$ to represent a deformation $\y$ of $S$,
\begin{equation}
\label{eq:y_representation}
\y=y_1\m+y_2\mper+y_3\e_3,
\end{equation}
where $y_i$ are smooth scalar fields on $S$, so that by \eqref{eq:connector}
\begin{equation}
\label{eq:nabla_y_representation_scalars}
\nabla\y=y_1\mper\otimes\cv+\m\otimes\nabla y_1-y_2\m\otimes\cv+\mper\otimes\nabla y_2+\e_3\otimes\nabla y_3.
\end{equation}
Letting, similarly, $\cv=c_1\m+c_2\mper$, we easily see that the vectors $\av$ and $\bv$ in \eqref{eq:nabla_y_representation} can be given the representation
\begin{subequations}\label{eq:a_b_representation}
\begin{eqnarray}
\av&=&\nay\m=(y_{1,1}-c_1y_2)\m+(y_{2,1}+c_1y_1)\mper+y_{3,1}\e_3,\\
\bv&=&\nay\mper=(y_{1,2}-c_2y_2)\m+(y_{2,2}+c_2y_1)\mper+y_{3,2}\e_3,
\end{eqnarray}
\end{subequations}
where we have used the expressions
\begin{subequations}\label{eq:nabla_y_i_representations}
	\begin{eqnarray}
	\nabla y_1&=&y_{1,1}\m+y_{1,2}\mper+y_{1,3}\e_3,\\
	\nabla y_2&=&y_{2,1}\m+y_{2,2}\mper+y_{2,3}\e_3,\\
	\nabla y_3&=&y_{3,1}\m+y_{3,2}\mper+y_{3,3}\e_3.
	\end{eqnarray}
\end{subequations}

Now, also in view of \eqref{eq:C}, we see that requiring $\y$ in \eqref{eq:y_representation} to be an isometric immersion satisfying \eqref{eq:C_stationary} reduces to enforcing the following conditions
\begin{equation}
\label{eq:isometry_conditions}
a^2=\lambda_1^2,\quad b^2=\lambda^2_2,\quad \av\cdot\bv=0.
\end{equation}
These, with the aid of \eqref{eq:a_b_representation}, read explicitly as
\begin{subequations}\label{eq:isometry_conditions_explicit}
	\begin{eqnarray}
	(y_{1,1}-c_1y_2)^2+(y_{2,1}+c_1y_1)^2+y_{3,1}^2&=&\lambda_1^2,\\
	(y_{1,2}-c_2y_2)^2+(y_{2,2}+c_2y_1)^2+y_{3,2}^2&=&\lambda_2^2,\\
	(y_{1,1}-c_1y_2)(y_{1,2}-c_2y_2)+(y_{2,1}+c_1y_1)(y_{2,2}+c_2y_1)+y_{3,1}y_{3,2}&=&0,
	\end{eqnarray}
\end{subequations}
which constitute a non-linear system of PDEs for the unknown functions $y_1$, $y_2$, and $y_3$.
As a consequence of Gauss' \emph{theorema egregium}, an isometric immersion characterized by \eqref{eq:C_stationary}
has Gaussian curvature dictated by $\m$ through the equation \cite{mostajeran:curvature,ozenda:blend}
\begin{equation}\label{eq:theorema_egregium_reduced}
K=\left(\lambda_1^2-\lambda_2^2\right)(c_2^2-c_1^2+c_{12}),
\end{equation}
where we have set $c_{12}=\m\cdot(\nabla\cv)\mper$.

Moreover, for a ridged isometry,  equations \eqref{eq:isometry_conditions_explicit} must be supplemented by the form appropriate to this setting of the jump condition in \eqref{eq:y_jump_condition}. Since both $\m$ and $\mper$ are continuous across any  plane curve $C$ (with unit tangent $\e$), by \eqref{eq:nabla_y_representation}, \eqref{eq:y_jump_condition} becomes
\begin{equation}
\label{eq:y_jump_condition_representation}
(\m\cdot\e)\jump{\av}+(\mper\cdot\e)\jump{\bv}=\zero.
\end{equation}
Letting $\e=\cos\chi\m+\sin\chi\mper$, since both $\y$ and $\cv$ are continuous across $C$, \eqref{eq:y_jump_condition_representation} reduces to the three scalar equations
	\begin{equation}\label{eq:y_jump_condition_representation_reduced}
	\jump{y_{1,1}}\cos\chi+\jump{y_{1,2}}\sin\chi=0,\quad
	\jump{y_{2,1}}\cos\chi+\jump{y_{2,2}}\sin\chi=0,\quad
	\jump{y_{3,1}}\cos\chi+\jump{y_{3,2}}\sin\chi=0.
	\end{equation}
While equations \eqref{eq:isometry_conditions_explicit} hold on the whole of $S$, despite the jumps that the gradients $\nabla y_1$, $\nabla y_2$, and $\nabla y_3$ may suffer across the curves $C_j$ that $\y$ transforms into the ridges $\curve_j$, equations \eqref{eq:y_jump_condition_representation_reduced} are valid only along such curves.

In the following section, we shall find solutions to \eqref{eq:isometry_conditions_explicit} and \eqref{eq:y_jump_condition_representation_reduced} in a special case. It will be expedient to compute on a ridge the inner product $\normal_1\cdot\normal_2$, which features in the expression for $f_r$ in \eqref{eq:ridge_energy_density}. To this end, we first recall  \eqref{eq:normal} and remark that 
\begin{equation}
\label{eq:normal_2}
\normal_2=(\av_1+\jump{\av})\times(\bv_1+\jump{\bv})=\normal_1+\av_1\times\jump{\bv}+\jump{\av}\times\bv_1,
\end{equation}
where use has also been made of \eqref{eq:y_jump_condition_representation}. Since $\av_1=\lambda_1\n_1$ and $\bv_1=\lambda_2\normal_1\times\n_1$, by \eqref{eq:isometry_conditions} it follows from \eqref{eq:normal_2} that 
\begin{equation}
\label{eq:normal_dot_normal}
\normal_1\cdot\normal_2=\frac{1}{\lambda_1^2}\av_1\cdot\av_2+\lambda_1^2\bv_1\cdot\bv_2-1,
\end{equation}
where we also employed \eqref{eq:lambda_1_2}. With $\av_i$ and $\bv_i$ given by \eqref{eq:a_b_representation}, we easily revert \eqref{eq:normal_dot_normal} into an expression featuring the traces of the gradient components $y_{i,j}$ on the two sides of the ridge under consideration.
 
\section{Ridged cones}\label{sec:hedgehog}
It is time now to put our theory to the test. In this section, we shall consider a classical example, already treated within the traditional theory \cite{modes:gaussian}, that of a disk $S$ of radius $R$ upon which the planar radial \emph{hedgehog} $\m$ has been imprinted. In polar coordinates $(\varrho,\vartheta)$, with associated orthonormal frame $(\ero,\et)$, $\m=\ero$ and $\mper=\et$. It is an easy exercise to check with the aid of \eqref{eq:connector} that then $c_1=0$, $c_2=\frac1\varrho$, and $c_{12}=-\frac{1}{\varrho^2}$, so that by \eqref{eq:theorema_egregium_reduced} $K=0$, independently of the prescribed principal stretches $\lambda_1$ and $\lambda_2$.

We shall use the representation \eqref{eq:y_representation} for $\y$ with
\begin{equation}
\label{eq:y_conical}
y_1=f\sin\psi\cos\varphi,\quad y_2=f\sin\psi\sin\varphi,\quad y_3=f\cos\psi,
\end{equation}
where $f$, $\psi$, and $\varphi$ are assumed to be picewise $C^2$-functions of $(\varrho,\vt)$. Thus $\psi$ and $\vp$ represents the polar and azimuthal angles of $\y$ in the movable frame $\framero$, while $f$ measures radial  dilation (or contraction). Some labour is required to see that with this choice the isometry conditions \eqref{eq:isometry_conditions_explicit} become
\begin{subequations}\label{eq:isometry_conditions_conical}
	\begin{eqnarray}
	f_{,\varrho}^2+f^2(\psi_{,\varrho}^2+\vp_{,\varrho}^2\sin^2\psi)&=&\lambda_1^2,\label{eq:isometry_conditions_conical_a}\\
	\frac{1}{\varrho^2}\left[f_{,\vt}^2+f^2\left(\psi_{,\vt}^2+(1+\vp_{,\vt})^2\sin\psi^2\right) \right]&=&\lambda_2^2,\label{eq:isometry_conditions_conical_b}\\
	\frac{1}{\varrho}\left\{f_{,\varrho}f_{,\vt}+f^2\left[\psi_{,\varrho}\psi_{,\vt}+\vp_{,\varrho}(1+\vp_{,\vt})\sin^2\psi\right]\right\}&=&0,\label{eq:isometry_conditions_conical_c}
	\end{eqnarray}
\end{subequations}
where commas denote partial derivatives in the variables $(\varrho,\vt)$. 

We shall look for solutions of \eqref{eq:isometry_conditions_conical} under the simplifying assumption that $f$ is a positive function of $\varrho$ only and both $\psi$ and $\vp$ are functions of $\vt$ only. 
The deformed surface $\surface$ is thus \emph{conical}, in the same class employed to represent crumpled sheets of paper in \cite{guven:paper,mueller:conical}.\footnote{This assumption is strongly motivated by the requirement that $K=0$, which demands that $\surface$ be developable. Strictly speaking, all these surfaces are singular at the tip, where all radii meet. This singularity could formally be removed by expunging  the centre of $S$. We rather prefer to keep it in place and tolerate the singularity it bears.} Further requiring that the centre of the disk $S$ is held fixed in a spontaneous deformation, we see that \eqref{eq:isometry_conditions_conical_a} has the unique (positive) solution
\begin{subequations}\label{eq:isometry_solutions}
\begin{equation}
\label{eq:isometry_condition_solution_a}
f(\varrho)=\lambda_1\varrho,
\end{equation}
and that \eqref{eq:isometry_conditions_conical_c} is identically satisfied, while \eqref{eq:isometry_conditions_conical_b} reduces to 
\begin{equation}
\label{eq:isomtery_condition_solution_b}
\psi'^2+(1+\vp')^2\sin^2\psi=\mu^2,
\end{equation}
where use has also been made of \eqref{eq:isometry_condition_solution_a} and we have set
\begin{equation}
\label{eq:mu_definition}
\mu:=\frac{\lambda_2}{\lambda_1}=\frac{1}{\lambda_1^2}.
\end{equation}
\end{subequations}
Hereafter in this section, a prime $'$ will denote differentiation with respect to $\vt$.\footnote{It is perhaps worth noting that, under the assumption that $f$ depends only on $\varrho$ and both $\psi$ and $\vp$ depend only on $\vt$, all three equations \eqref{eq:isometry_solutions} also follow from simply requiring that $\det\C=1$.}

We now focus on solving \eqref{eq:isomtery_condition_solution_b} subject to the periodic boundary conditions
\begin{equation}
\label{eq:boundary_conditions}
\psi(0)=\psi(2\pi),\quad\vp(0)=\vp(2\pi).
\end{equation}
For a given function $\psi$ satisfying the first equality in \eqref{eq:boundary_conditions} and such that $\sin\psi\neq0$ and $\psi'^2\leqq\mu^2$, \eqref{eq:isomtery_condition_solution_b} is solved by integrating 
\begin{equation}
\label{eq:phi_prime}
\vp'=-1+\frac{\sqrt{\mu^2-\psi'^2}}{\sin\psi},
\end{equation}
provided that the following condition is met,
\begin{equation}
\label{eq:I_functional}
I:=\int_0^{2\pi}\frac{\sqrt{\mu^2-\psi'^2}}{\sin\psi}\dd\vt=2\pi,
\end{equation}
for the second equality in \eqref{eq:boundary_conditions} to be valid too. Clearly, for $\mu\leqq1$, \eqref{eq:I_functional} is solved by $\psi\equiv\arcsin\mu$, which corresponds to a circular cone when $\mu<1$. 

It is for $\mu>1$ where our discrete ridge model comes to play.
We look for ridged isometric immersions representable in the class \eqref{eq:y_conical}. First, we see how to write the jump conditions \eqref{eq:y_jump_condition_representation_reduced} in the present context. Here $\chi=0$, as jumps may only occur along radii of the disk $S$. Moreover, by \eqref{eq:isometry_condition_solution_a}, equations \eqref{eq:y_conical} imply that $\jump{y_{1,1}}=\jump{y_{2,1}}=\jump{y_{3,1}}=0$, independently of $\psi$, so that \eqref{eq:y_jump_condition_representation_reduced} is identically satisfied. We shall thus look for solutions $\psi$ of \eqref{eq:isomtery_condition_solution_b} that are piecewise of class $C^1$. Since \eqref{eq:isomtery_condition_solution_b} must be valid on the whole of $S$, a jump in $\psi'$ is admissible only if
\begin{equation}
\label{eq:jump_psi_prime_squared}
\jump{\psi'^2}=0,
\end{equation}
while, by \eqref{eq:phi_prime}, no jump in $\vp'$ is allowed. 

It is a simple matter to show that for a conical surface described by \eqref{eq:y_conical} the vectors $\av$ and $\bv$ in \eqref{eq:a_b_representation} can be given the following expressions (also by use of \eqref{eq:isometry_condition_solution_a}),
\begin{subequations}\label{eq:a_b_conical_expressions}
	\begin{eqnarray}
	\av&=&\lambda_1\left(\sin\psi\cos\vp\ero+\sin\psi\sin\vp\et+\cos\psi\e_3\right),\\
	\bv&=&\lambda_1\{[\psi'\cos\psi\cos\vp-(1+\vp')\sin\psi\sin\vp]\ero
	+[\psi'\cos\psi\sin\vp+(1+\vp')\sin\psi\cos\vp]\et\nonumber\\&-&\psi'\sin\psi\e_3\}.
	\end{eqnarray}
\end{subequations} 
It follows from these equations and \eqref{eq:normal_dot_normal} that on a ridge
\begin{equation}
\label{eq:conical_normal_dot_normal}
\normal_1\cdot\normal_2=1-2\frac{\psi'^2}{\mu^2},
\end{equation}
where use has also been made of both \eqref{eq:isomtery_condition_solution_b} and \eqref{eq:jump_psi_prime_squared}. Moreover, since $\tangentper\cdot\n=0$, by \eqref{eq:ridge_energy_density} the ridged energy density (scaled to $\frac83h^2$) simply reduces to $f_r=\arccos^2(\normal_1\cdot\normal_2)$.

We must still enforce \eqref{eq:boundary_conditions} and \eqref{eq:I_functional}. We do so by means of a geometric construction, which is fully substantiated in Appendix~\ref{sec:construction}. We take full advantage of our unconventional approach that here renounces to approximate the smooth deformed surface $\surface$ with a large (albeit finite) number of ridges. Granting (up-down) symmetry to $\surface$, we shall be contented with capturing the simplest feature: how many (up ad down) \emph{folds} it possesses. Thus, here the unknown number of folds will just be the number of admitted ridges. Minimizing the total ridge energy will provide the optimal number of folds.

\begin{figure}[h]
	\centering
	\begin{subfigure}[t]{0.3\textwidth}
		\centering
		\includegraphics[width=\textwidth]{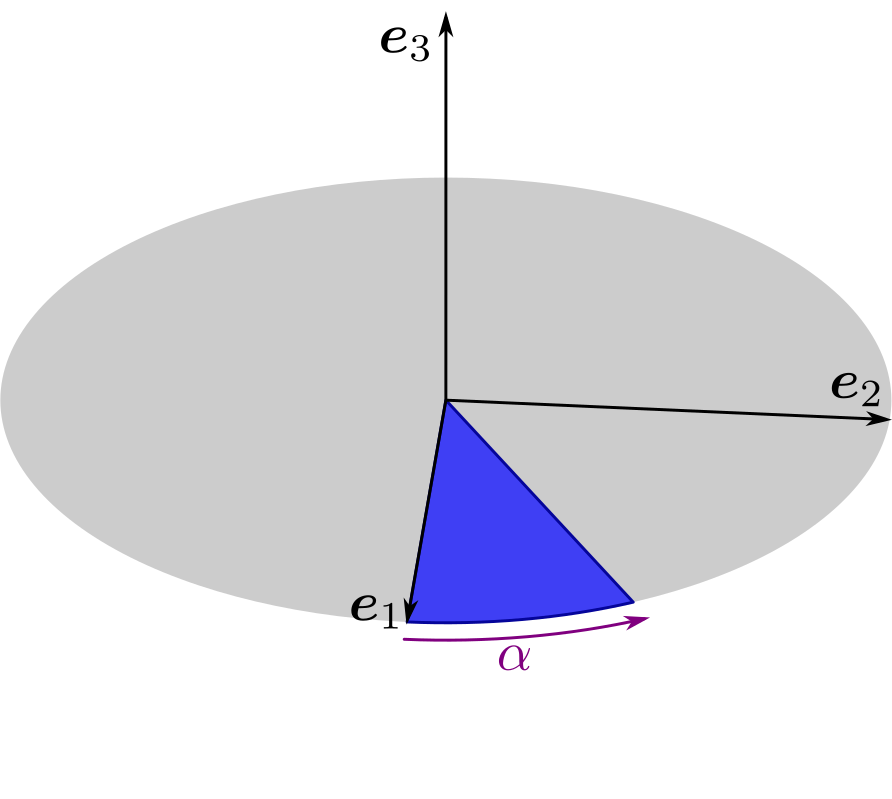}
		\caption{Step 1: Choose an integer $n\geq\mu$ and select a reference sector with amplitude $\alpha=\frac{\pi}{2n}$. Here $\mu=1.7$ and $n=3$.}
		\label{fig:step01}
	\end{subfigure}
	$\quad$
	\begin{subfigure}[t]{0.3\textwidth}
		\centering
		\includegraphics[width=\textwidth]{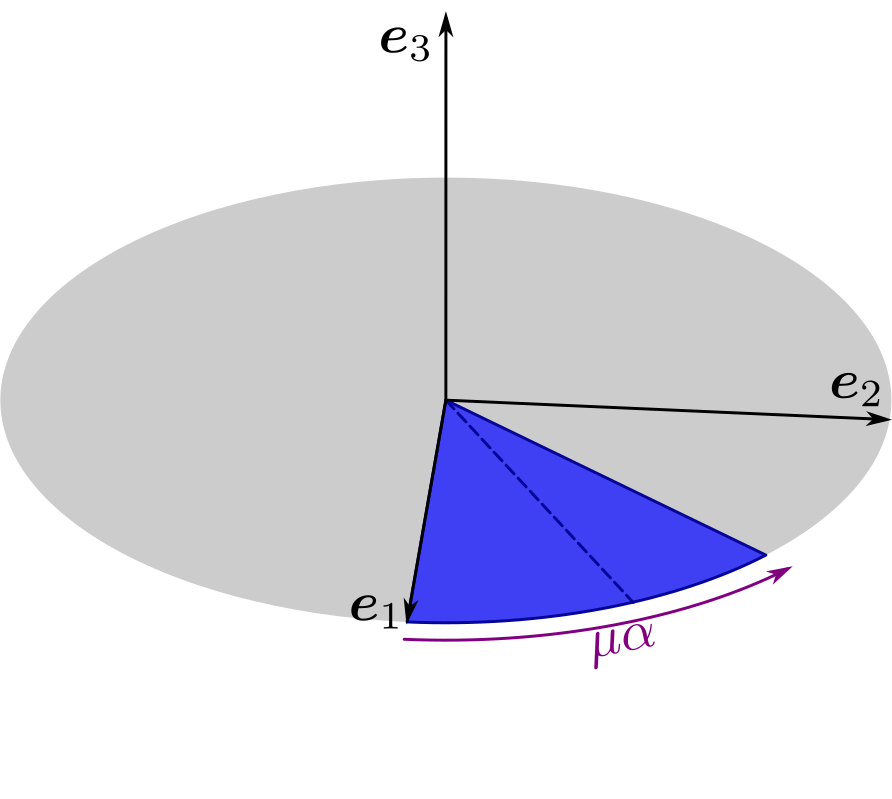}
		\caption{Step 2: Stretch uniformly the reference sector into one with amplitude $\mu\alpha$.}
		\label{fig:step02}
	\end{subfigure}
	$\quad$
	\begin{subfigure}[t]{0.3\textwidth}
		\centering
		\includegraphics[width=\textwidth]{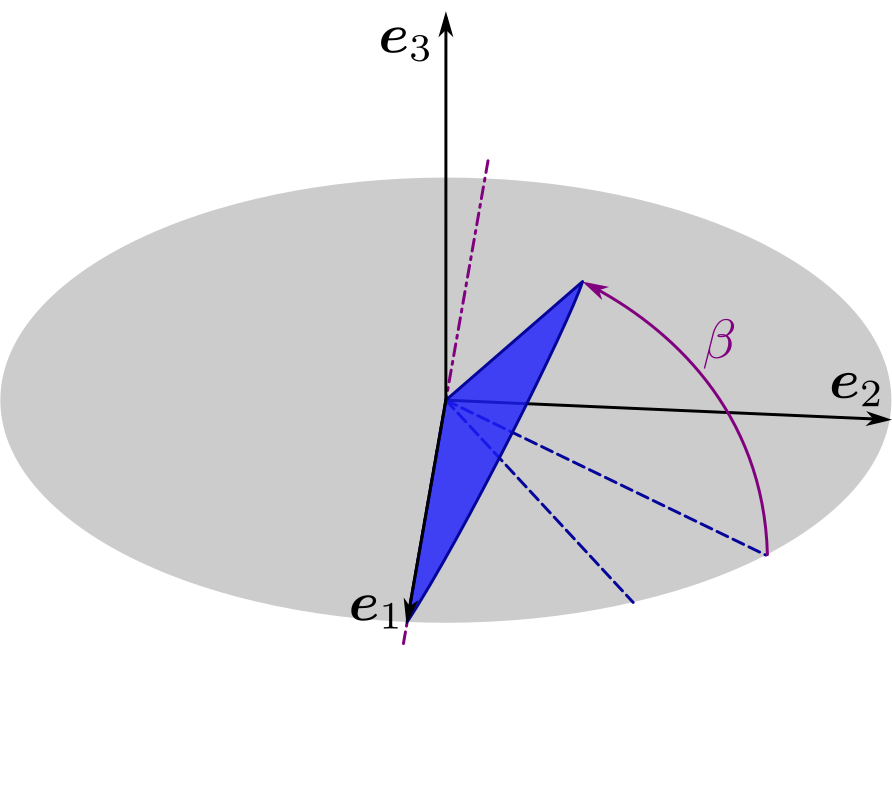}
		\caption{Step 3: Rotate around $\bm{e}_{1}$ by the angle $\beta$.}
		\label{fig:step03}
	\end{subfigure}
	 \begin{subfigure}[t]{0.3\textwidth}
		\centering
		\includegraphics[width=\textwidth]{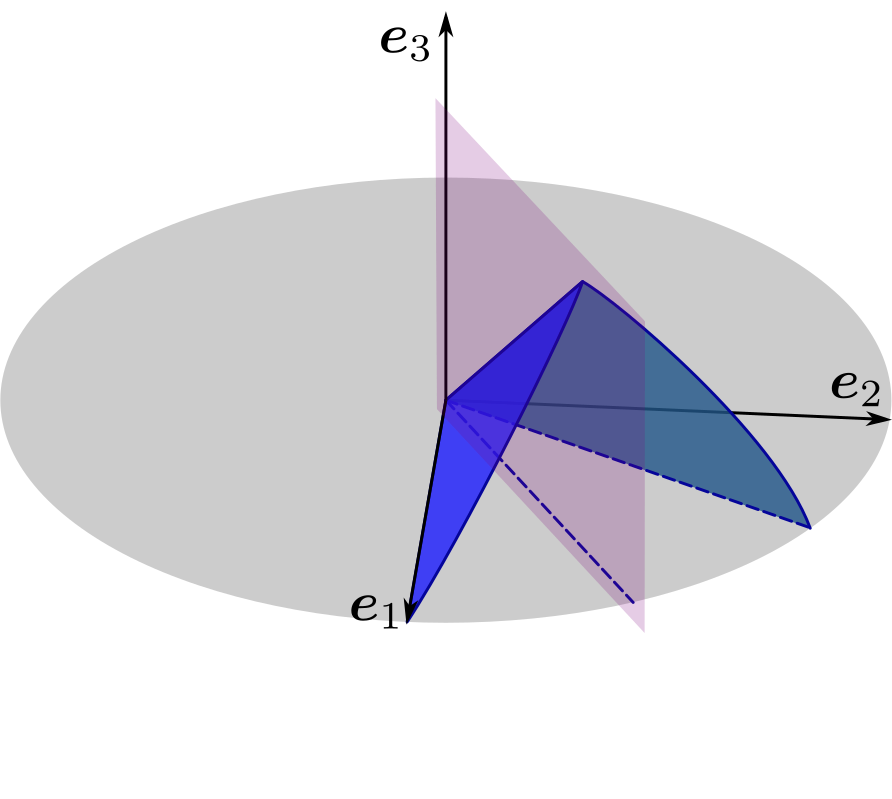}
		\caption{Step 4: Reflect across the vertical plane containing the lifted edge.}
		\label{fig:step07}
	\end{subfigure}
	$\quad$
	\begin{subfigure}[t]{0.3\textwidth}
		\centering
		\includegraphics[width=\textwidth]{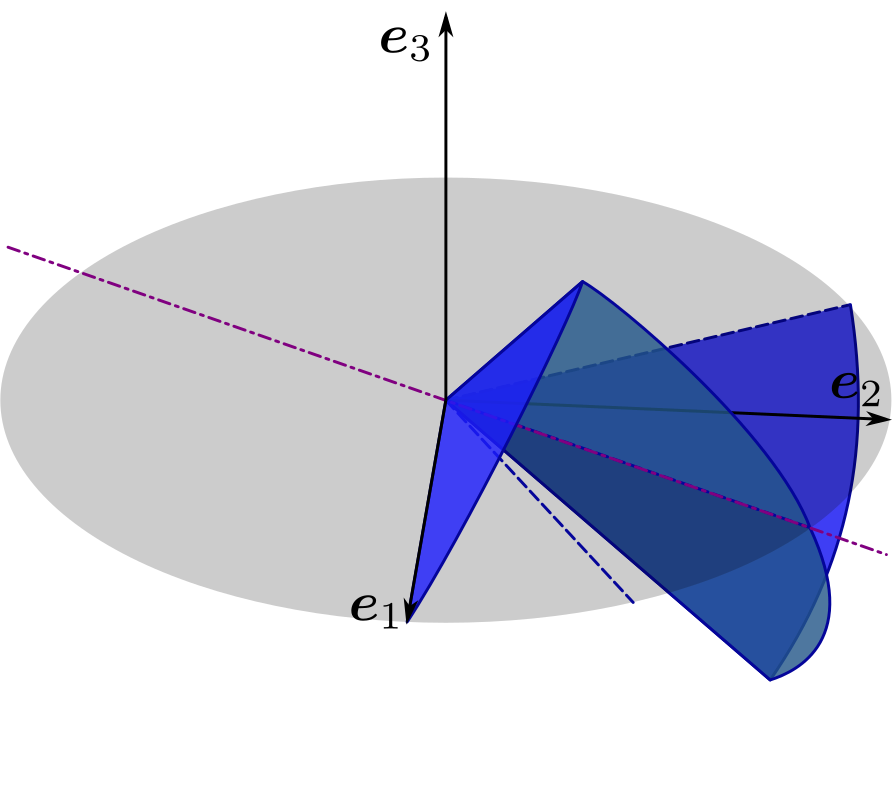}
		\caption{Step 5: Rotate by $\pi$ around the edge lying on $S$.}
		\label{fig:step08}
	\end{subfigure}
	$\quad$
	\begin{subfigure}[t]{0.3\textwidth}
		\centering
		\includegraphics[width=\textwidth]{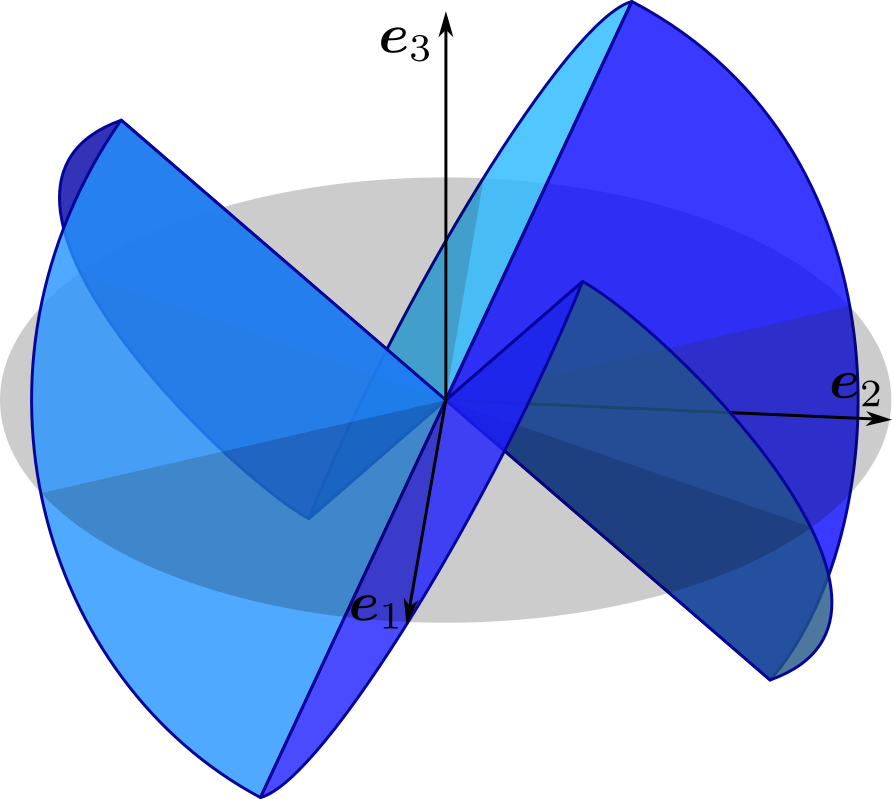}
		\caption{Step 6: Repeat $n-1$ times.}
		\label{fig:step09}
	\end{subfigure}
	\caption{Geometric construction of a ridged immersion of the unit disk $S$.}
	\label{fig:construction}
\end{figure}

By rescaling lengths, we can assume that $S$ is the unit disk, stretched circumferentially by $\mu$ into a symmetric ridged immersion through the following steps (see Fig.~\ref{fig:construction}).
\begin{inparaenum}[(1)]
	\item For a given $\mu>1$, choose an integer $n\geqq\mu$ and take a circular sector of $S$ of amplitude $\alpha:=\frac{\pi}{2n}$ (the reference sector). \label{step:1}
	\item Stretch it so that its amplitude becomes $\mu\alpha$. \label{step:2}
	\item Rotate the stretched sector around one of its edges by an appropriately chosen angle $\beta$. \label{step:3}
	\item Reflect the rotated sector across the vertical plane containing the lifted edge. \label{step:4}
	\item Rotate by $\pi$ the reflected sector around its edge lying on $S$. \label{step:5}
	\item Repeat $n-1$ times the preceding steps. \label{step:6}
\end{inparaenum}	

This construction works, making sure that the generated surface closes on itself, if $\beta$ is such that the reflection plane in step \eqref{step:4} cuts $S$ where lied the edge of the reference sector (see Fig.~\ref{fig:step07}), that is, if (see Appendix~\ref{sec:construction}).
\begin{equation}
\label{eq:beta}
\beta:=
\begin{cases}
\arccos\frac{\tan\alpha}{\tan(\mu\alpha)} &\text{if }\mu<n,\\
\frac{\pi}{2} &\text{if }\mu=n.
\end{cases}
\end{equation}	
A continuous, piecewise differentiable $4\alpha$-periodic function $\psi$ is obtained in \eqref{eq:psi_extended} by extending over $\real$ the function $\psi^\ast$ defined on $[0,\alpha]$ by
\begin{equation}
\label{eq:psi_star}
\psi^\ast(\vt):=\arccos(\sin\beta\sin(\mu\vt)).
\end{equation}
Correspondingly,  the function $\vp^\ast$ associated with $\psi^\ast$ is given in $[0,\alpha]$  by  (see again Appendix~\ref{sec:construction})
\begin{equation}
\label{eq:phi_star}
\vp^\ast(\vt):=-\vt+\arccos\frac{\cos(\mu\vt)}{\sin\psi^\ast(\vt)}.
\end{equation}
Its $C^1$-extension $\vp$  over $\real$ compatible with \eqref{eq:phi_prime} is recorded in \eqref{eq:phi_extended}.
It is  a boring, but simple exercise to check that functions $\psi^\ast$ and $\vp^\ast$ satisfy \eqref{eq:isomtery_condition_solution_b} identically, and so do their extensions $\psi$ and $\vp$.

\begin{figure}[h]
	\centering
	\begin{subfigure}[t]{.8\textwidth}
		\centering
		\includegraphics[width=0.4\textwidth]{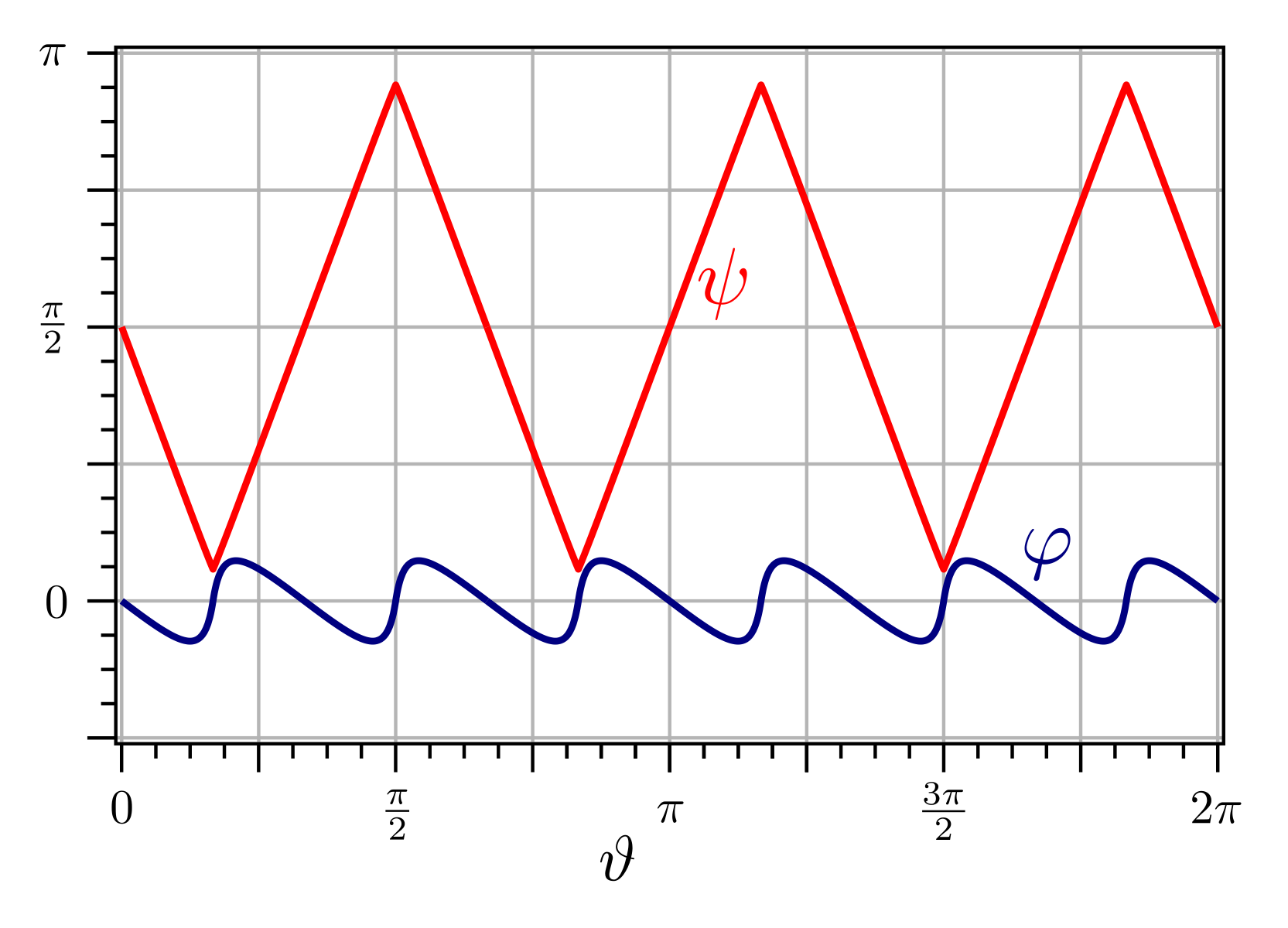}\
		\includegraphics[width=0.3\textwidth]{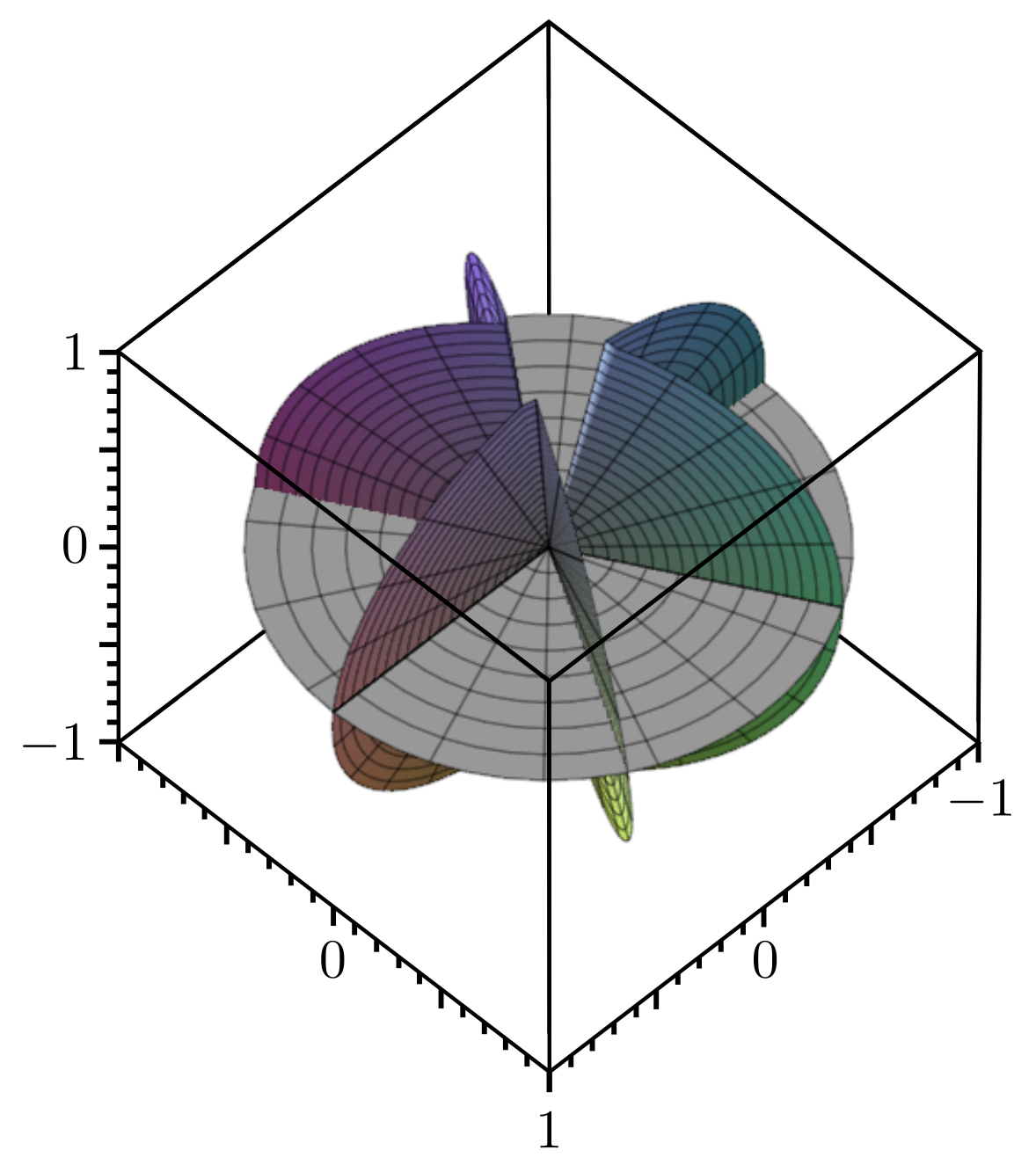}
		\caption{$\mu=2.7$ and $n=3$.}
	\end{subfigure}
	\begin{subfigure}[t]{.8\textwidth}
		\centering
		\includegraphics[width=0.4\textwidth]{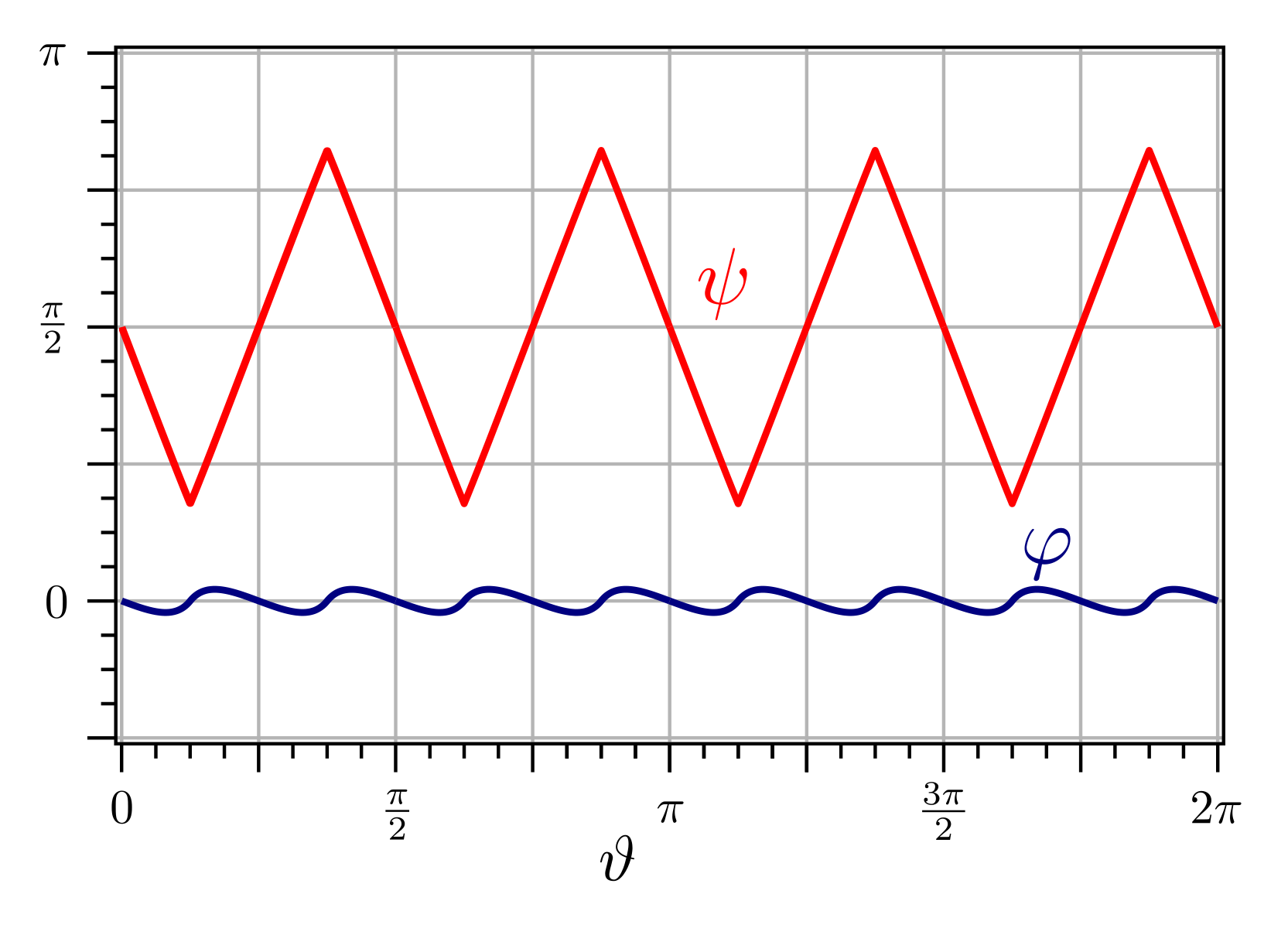}\
		\includegraphics[width=0.3\textwidth]{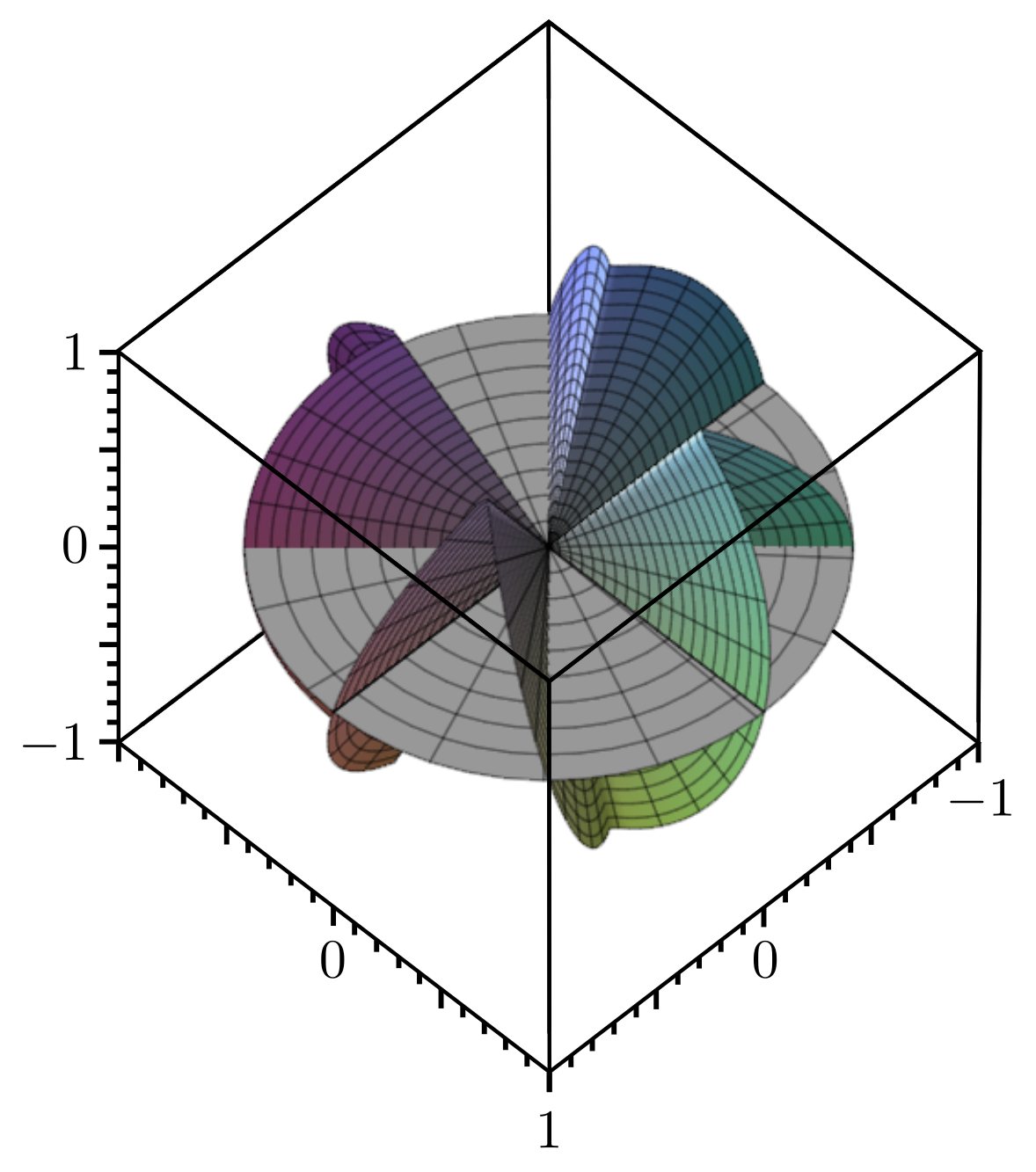}
		\caption{$\mu=2.7$ and $n=4$.}
	\end{subfigure}
	\begin{subfigure}[t]{.8\textwidth}
		\centering
		\includegraphics[width=0.4\textwidth]{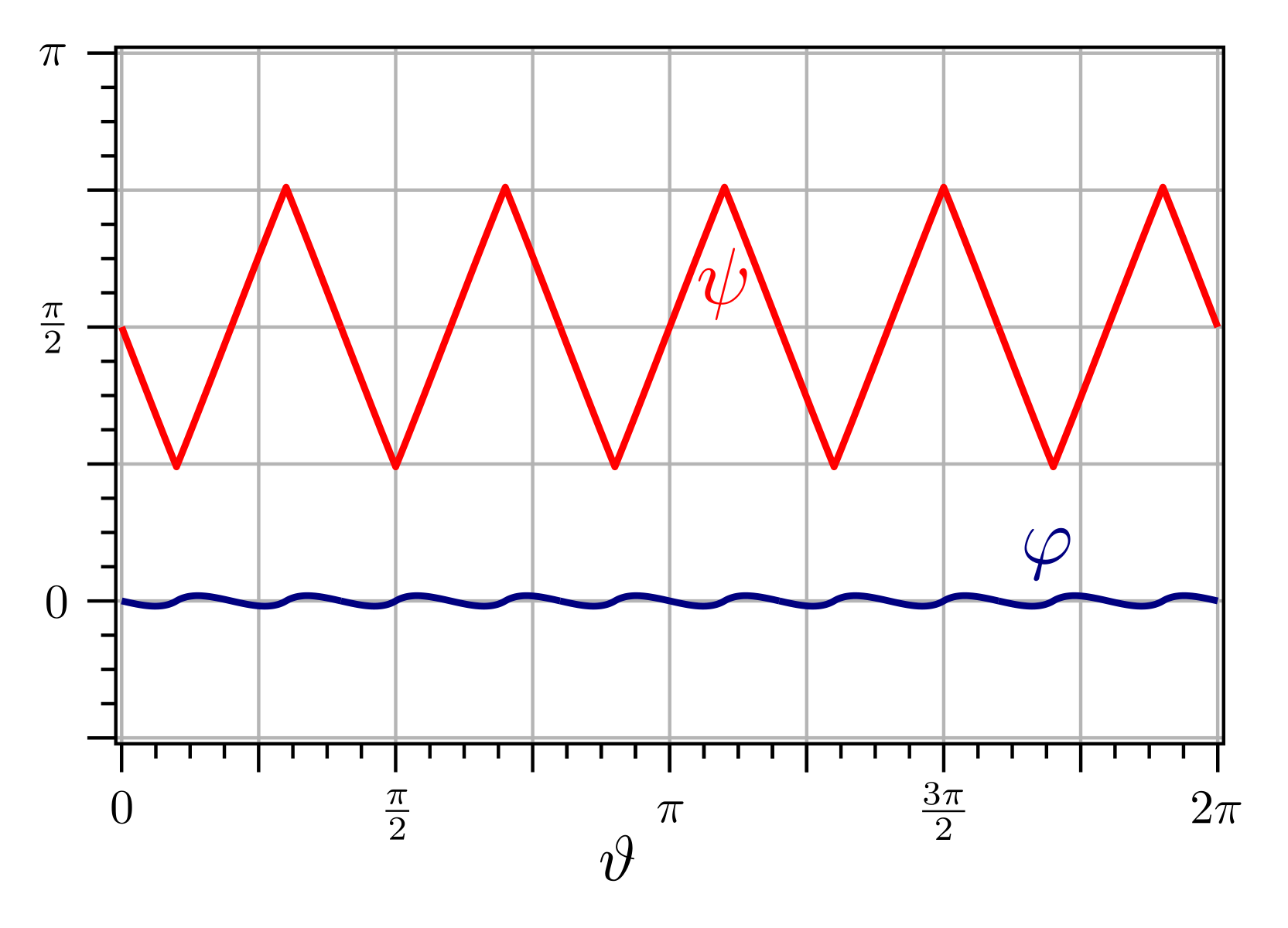}\
		\includegraphics[width=0.3\textwidth]{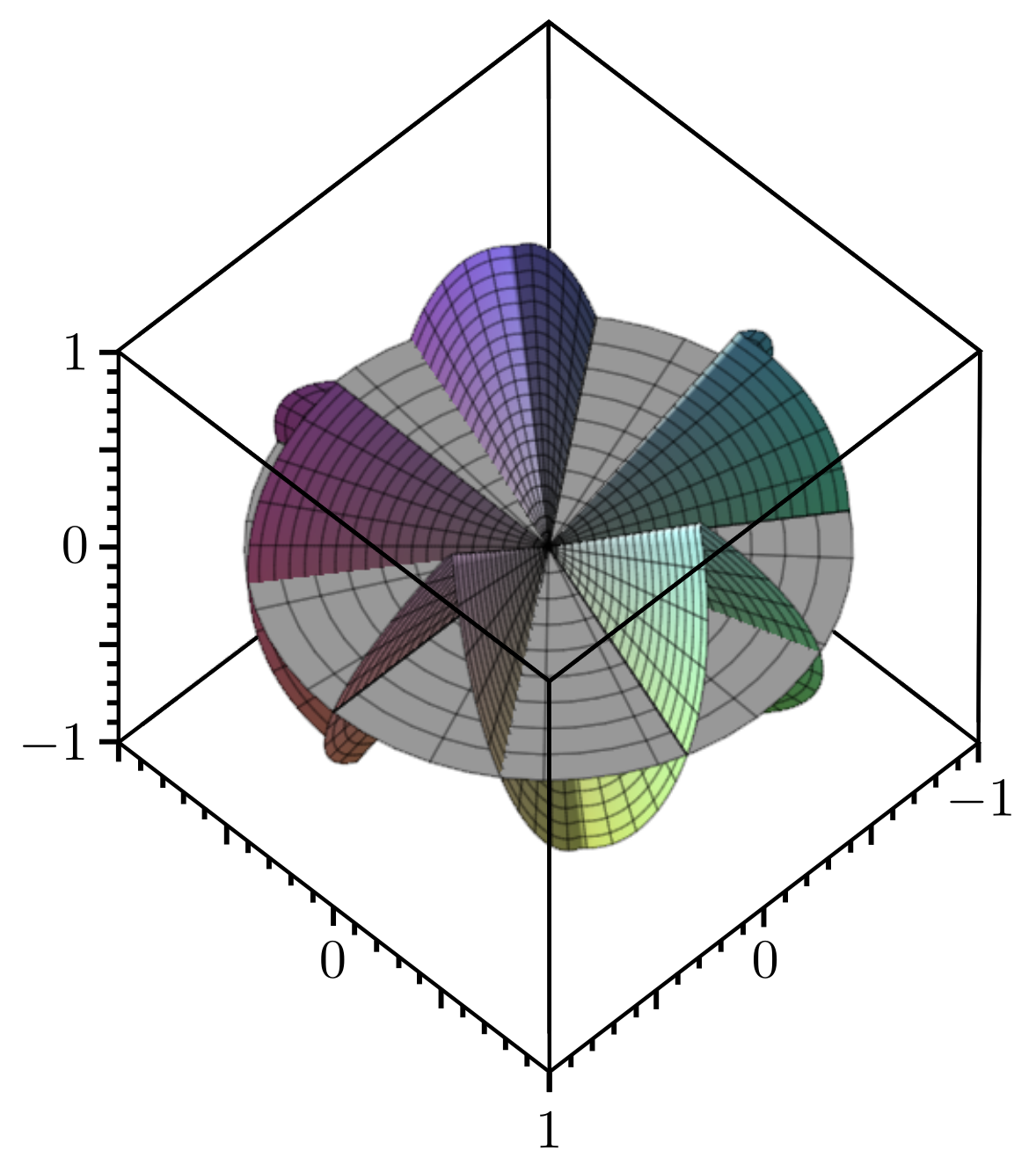}
		\caption{$\mu=2.7$ and $n=5$.}
	\end{subfigure}
	\caption{On the left: Functions $\psi$ (red) and $\varphi$ (blue) for $\mu=2.7$ and different values of $n$. On the right:  Corresponding ridged immersions of the unit disk $S$.}
	\label{fig:psi_phi_shapes}
\end{figure}

\begin{figure}[h]
		\centering
		\includegraphics[width=0.32\textwidth]{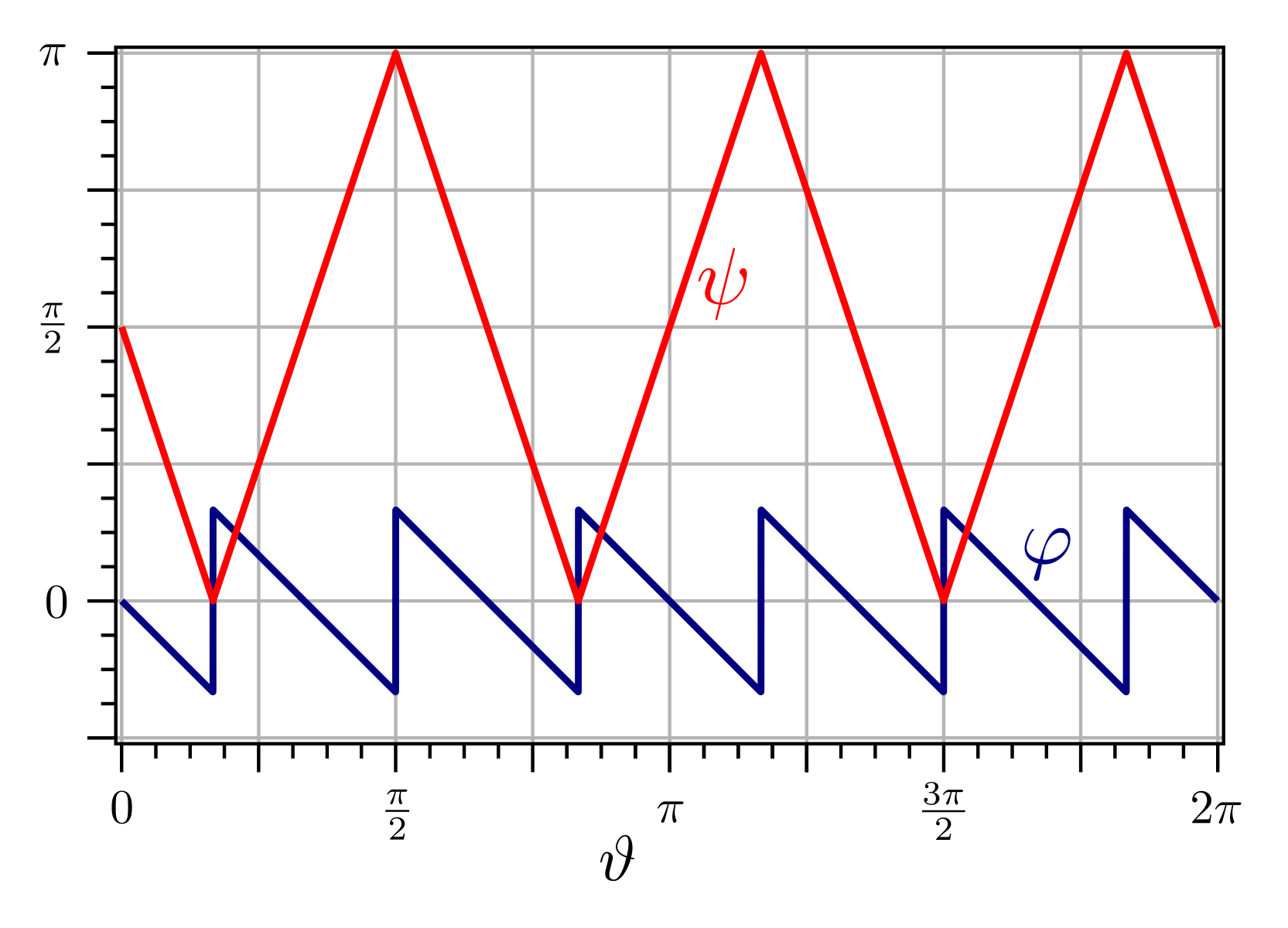}\
		\includegraphics[width=0.24\textwidth]{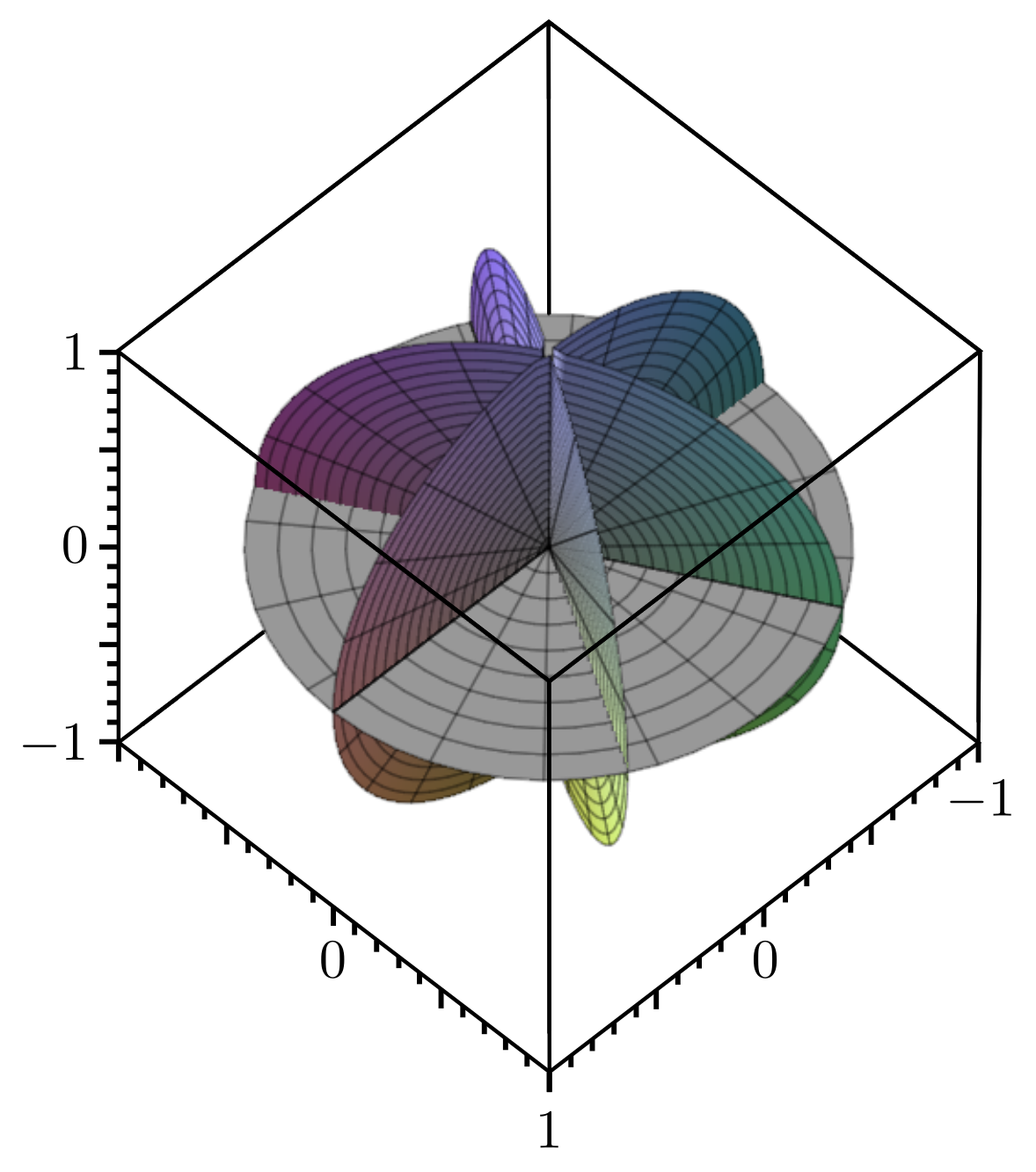}
	\caption{When the geometric construction of a ridged immersion of $S$ becomes singular. On the left: Functions $\psi$ (red) and $\varphi$ (blue) for $\mu=n=3$. On the right: Corresponding ridged immersion of $S$ with all ridges collapsed on the vertical axis of $S$.}
	\label{fig:mu_3}
\end{figure}

Figure~\ref{fig:psi_phi_shapes} shows examples of the functions $\psi$ and $\vp$ so generated alongside with the conical surface produced by the corresponding ridged immersions of the unit disk $S$.
It should be noted that whenever $\mu=n$ this construction becomes singular, though it is still applicable. In such a case, all ridges are degenerate and lie on the vertical axis $\e_3$ of the disk $S$; all faces of $\surface$ are vertical as well and $\vp$ becomes discontinuous (as shown in Fig.~\ref{fig:mu_3} for $\mu=3$).

\begin{figure}[h]
	\centering
	\includegraphics[width=0.5\textwidth]{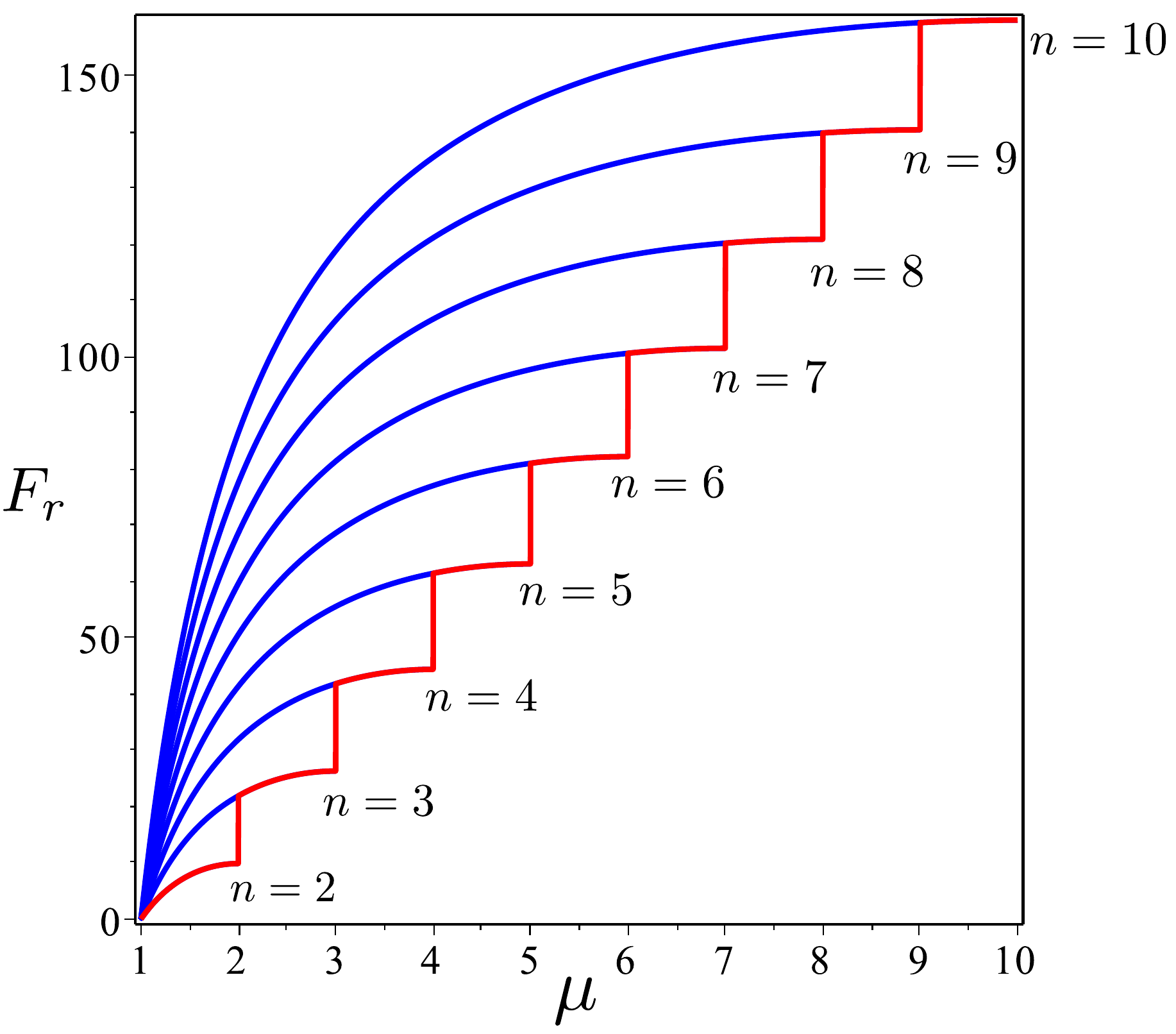}
	\caption{The function $F_r$ in \eqref{eq:F_r} is plotted against $\mu$ in the intervals $[1,n]$, for several values of $n$ (blue graphs). The lower envelope (red graph) is the plot of $F_r(\lceil\mu\rceil,\mu)$ against $\mu$.}
	\label{fig:F_r_plot}
\end{figure}

The above construction produces $N=2n$ ridges, at each of which $\psi'^2=\psi'^2(\alpha)$, so that by \eqref{eq:conical_normal_dot_normal} the total (dimensionless) ridge energy $F_r$ is (see \eqref{eq:ridge_energy_multiple} and \eqref{eq:periodic_values})
\begin{equation}
\label{eq:F_r}
F_r(n,\mu)=2n\arccos^2\left(1-\frac{2}{\sin^2\left(\frac{\mu\pi}{2n}\right)}\left[\cos^2\left(\frac{\pi}{2n}\right)-\cos^2\left(\frac{\mu\pi}{2n}\right) \right]\right).
\end{equation}
Plots of $F_r$ against $\mu$ for several values of $n$ are depicted in Fig.~\ref{fig:F_r_plot}. They show that, for a given $\mu$, $F_r$ is
minimized for $n=\lceil\mu\rceil$, when the number of ridges is the least possible (as was perhaps to be expected).\footnote{By $\lceil\mu\rceil$, we mean the smallest integer greater than or equal to $\mu$.}

\section{Conclusions}\label{sec:conclusions}
Common wisdom has it that in sufficiently thin sheets of nematic polymer networks, as in all elastic material for that matter (see, for example, \cite[p.\,396]{rayleigh:theory} or \cite[p.\,404]{audoly:geometry}), the bending energy (which scales as the cube of the thickness) may be neglected relative to the stretching energy (which scales linearly in the thickness). When activated, a nematic polymer network suffers a spontaneous deformation that attempts to transfer on the current shape the metric tensor that minimizes the stretching energy, which (with a slight abuse of language) we called an isometric immersion, for short. Such an immersion would generally depend on the nematic director imprinted on the sheet at the time of crosslinking, and, as is well known, it may fail to exist.

We started from relaxing  the requirement of smoothness for an isometric immersion, thus removing a possible obstacle to its existence. We allowed for \emph{ridges} in the immersed surfaces $\surface$ representing deformed sheets; these are lines where the normal to $\surface$ suffers a jump. Clearly, ridges do not come for free (nothing does). If they did, we would be overwhelmed with a superabundance of  shapes, for which we would lack a selecting energy criterion (as all would have the same stretching energy). 

We thought of ridges as concentrations of bending energy; we put forward a model to compute the energy they bear distributed along their length. To accomplish this task, we employed a formula for the bending energy recently derived from the ``trace formula'' valid in three space dimensions \cite{ozenda:blend}.

We showed that the ridge energy density (per unit length) $f_r$ scales quadratically with the sheet's thickness, and so it represents a contribution intermediate between stretching and bending energies. The formula we obtained for $f_r$ not only depends (symmetrically) on the normals to the adjoining sides (as was perhaps to be expected), but also on the orientation of the nematic director relative to the tangent to the ridge.

We applied our theory to the case where a planar hedgehog is imprinted on a flat disk at the time of crosslinking. We studied the total elastic energy, including the new ridge energy, in a class of conical deformations not new in the literature. In the regime where the radii of the reference disk shrink and the circumferences expand, by identifying the ridges of the discrete model with the folds of the continuum model, we used the total ridge energy as a selection criterion to determine the optimal number of folds.

The ridged cones that we found as energy minimizers are neither the \emph{developable cones} of \cite{lobkovsky:scaling,ben_amar:crumpled,cerda:conical,cerda:elements,cerda:confined,guven:paper} nor the \emph{excess cones} of \cite{mueller:conical}, as they do not share the degree of  smoothness that both the latter and the former have in common. But we trust that, under similar circumstances, they have in common the same number of folds. 

Of course, we cannot expect that a rubber like material will spontaneously take on sharp ridges when activated by a change in its internal material organization. Our model, as applied here, has more the flavour of a vicarious theory, where a distributed bending energy is replaced by one concentrated at a number of places. As we have shown, it is, nevertheless, predictive, at least of the expected number of folds. Since $F_r$ in \eqref{eq:F_r} is minimized for $n=\lceil\mu\rceil$ and, correspondingly, the optimal number of ridges/folds is $N=2\lceil\mu\rceil$, it follows from \eqref{eq:mu_definition} and \eqref{eq:lambda_1_2} that
\begin{equation}
\label{eq:number_of_ridges}
N=2\left\lceil\sqrt{\frac{s_0+1}{s+1}}\right\rceil\geqq4\quad\text{for}\quad s<s_0,
\end{equation} 
where $s$ is the activation parameter of the theory.

We expect that  \eqref{eq:number_of_ridges} would reproduce the number of folds predicted by an elastic theory based on the full blown energy, where stretching and bending components are blended together and compete on different length scales. 

Symmetry and identification of folds with ridges played a role in deriving \eqref{eq:number_of_ridges}.  The ridge energy obtained in this paper, like the notion itself of ridged immersions, is susceptible of further applications, if we relax the ridge/fold identification and take the more traditional approach of considering our (simplified) discrete model as an approximation of the (more difficult) continuum model (with bending energy). This approximation, which is expected to improve upon increasing the number of ridges, should be justified by a convergence assessment. For a finite (but large) number of ridges unrestricted by symmetry requirements, it would provide a good test for the number of folds predicted by \eqref{eq:number_of_ridges}. More generally, an appropriate decomposition of the reference surface $S$ could be ridge-immersed in a triangulation of the deformed surface $\surface$ to determine the optimal shape of the activated film. These extensions are presently being studied.

\begin{acknowledgments}
Some of the contents of this paper were first illustrated by E.G.V. in a lecture given in December 2019  at the Institute for Computational and Experimental Research in Mathematics (ICERM) in Providence, RI, during the Workshop on \href{https://icerm.brown.edu/video_archive/?play=2112}{\emph{Numerical Methods and New Perspectives for Extended Liquid Crystalline Systems}}. The kindness of the organizers of the Workshop and the generous hospitality of ICERM are gratefully acknowledged. 
The work of A.P. was supported financially by the Department of Mathematics of the University of Pavia as part of the activities funded by the Italian MIUR under the nationwide Program ``Dipartimenti di Eccellenza (2018-2022).''
We are indebted to an anonymous critical reviewer of an earlier version of this manuscript for suggesting, indirectly, to test our approach against the golden standard of the \emph{elastica} theory.
\end{acknowledgments}

\section*{Authors contributions}
Both authors were involved in the preparation of the manuscript.
They have read and approved the final manuscript.

\appendix
\section{Geometric Construction}\label{sec:construction}
In this appendix, we provide further details on the geometric construction employed in Sect.~\ref{sec:hedgehog} to produce a ridged isometric immersion of the unit disk $S$. In particular, our objective will be to justify both the formula for $\beta$ in \eqref{eq:beta} and that for $\psi^\ast$ in \eqref{eq:psi_star}.

\begin{figure}[h]
\centering
\begin{subfigure}[t]{0.3\textwidth}
	\centering
	\includegraphics[width=\textwidth]{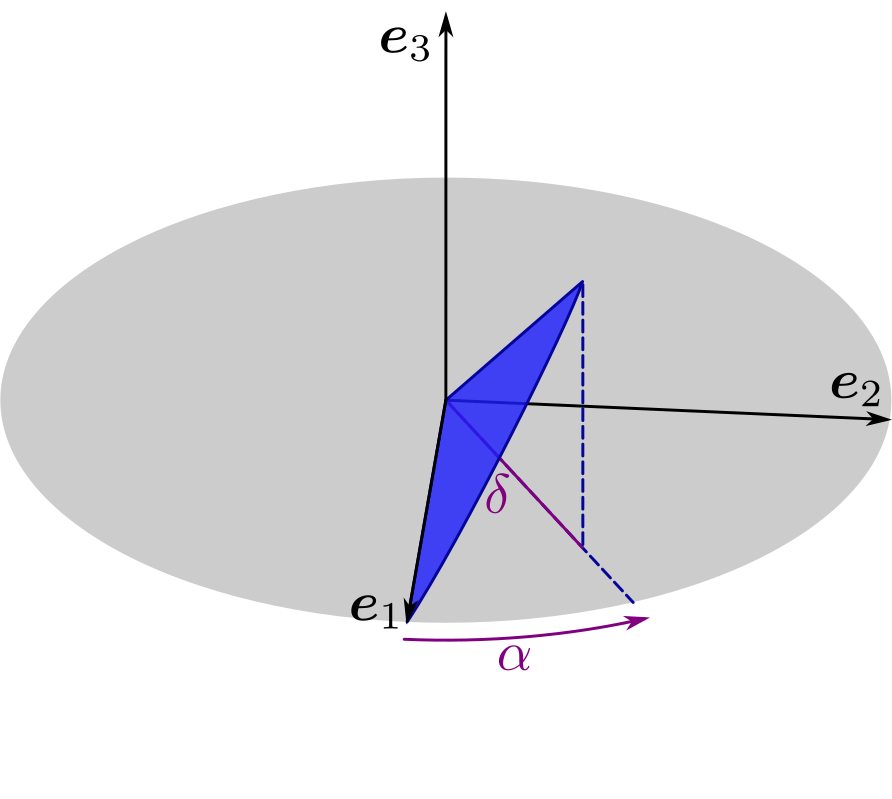}
	\caption{One edge of the first immersed sector is $\bm{e}_{1}$, while the projection of the other edge on $S$ has length $\delta$ and azimuthal angle $\alpha$.}
	\label{fig:step04}
\end{subfigure}
$\quad$
\begin{subfigure}[t]{0.3\textwidth}
	\centering
	\includegraphics[width=\textwidth]{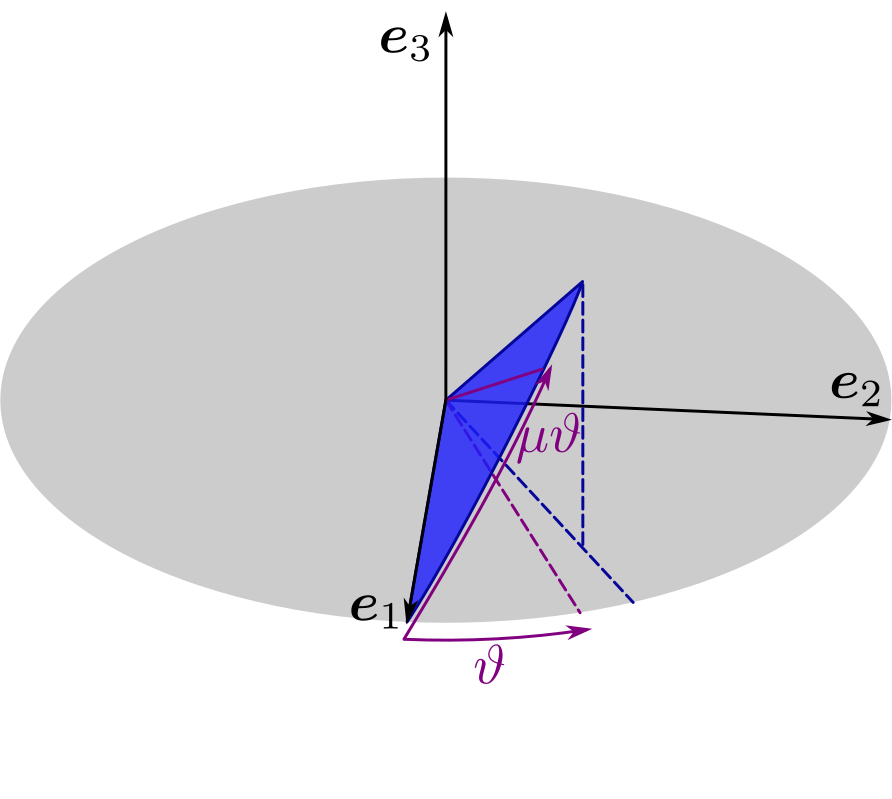}
	\caption{The ray that makes the angle   $\vartheta$ with $\e_{1}$ in the reference sector is mapped into the ray that makes the angle $\mu\vartheta$ with $\e_{1}$ in the first immersed sector.}
	\label{fig:step05}
\end{subfigure}
$\quad$
\begin{subfigure}[t]{0.3\textwidth}
	\centering
	\includegraphics[width=\textwidth]{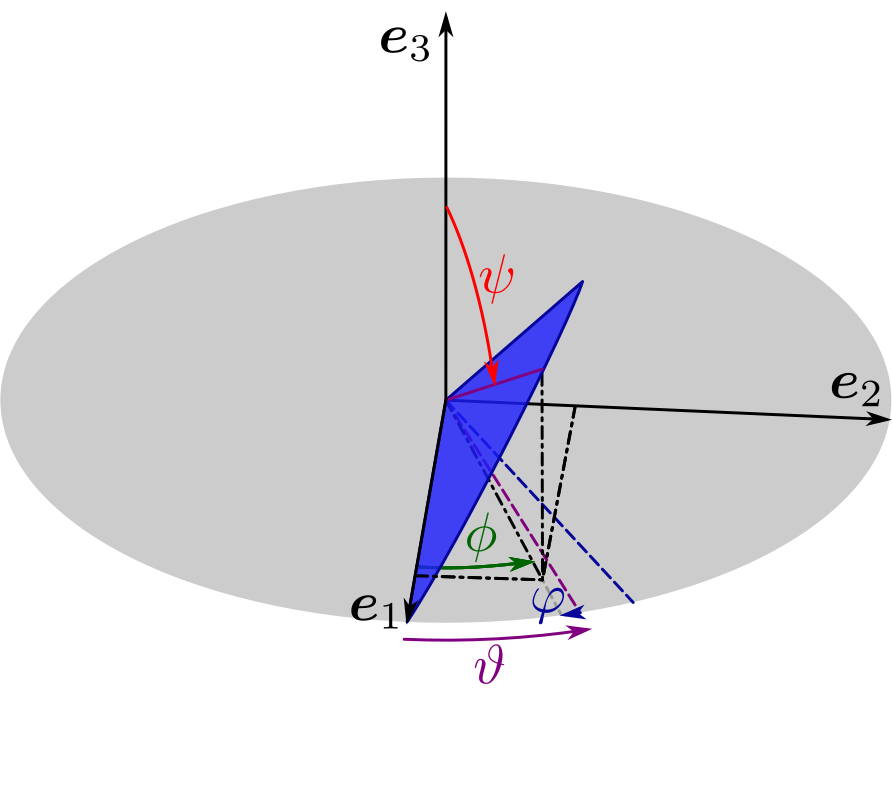}
	\caption{The ray that makes the angle  $\vartheta$ with $\e_{1}$ in the reference sector is mapped into the ray with azimuthal angle $\phi=\vt+\varphi$ and polar angle $\psi$. Here $\varphi<0$.}
	\label{fig:step06}
\end{subfigure}
\caption{Geometric details for the first immersed sector.}
\label{fig:properties}
\end{figure}

Take  a circular sector of $S$ with amplitude $\alpha$ (the reference sector) and  stretch it uniformly to obtain a sector with amplitude $\mu\alpha$. Then rotate the stretched sector around $\bm{e}_{1}$ by $\beta$ and reflect the rotated sector across the vertical plane passing through the unit vector $\e(\alpha):=\cos\alpha\e_1+\sin\alpha\e_2$. Finally, rotate by $\pi$ the reflected sector around $\e(2\alpha)$. Thus, we have ridged-immersed a sector with amplitude $4\alpha$. By replicating $n-1$ times this immersion, we finally  obtain the entire surface $\surface$. Figure~\ref{fig:properties} shows the geometric details of the first immersed sector (with amplitude $\alpha$).

More precisely, for all $\vartheta\in[0,\alpha]$ we define the mapping 
\begin{equation}\label{eq:E_mapping_definition}
\e(\vt)\mapsto\E(\vt):=\R_{\e_{1},\beta}\e(\mu\vartheta),
\end{equation}
where $\R_{\e_{1},\beta}$ is the rotation around $\vc{e}_{1}$ by the angle $\beta$. Explicitly, $\E(\vt)$ is given by
\begin{eqnarray}\label{eq:rotation}
\E(\vt)&=&\left(\I+\sin\beta\W(\e_1)-(1-\cos\beta)\proj(\e_1)\right)\e(\vt)\nonumber\\
&=& \cos(\mu\vartheta)\vc{e}_{1}+\sin(\mu\vartheta)\cos\beta\vc{e}_{2}+\sin(\mu\vartheta)\sin\beta\vc{e}_{3},
\end{eqnarray}
where $\proj(\e_1)$ denotes the projection onto the plane orthogonal to $\e_1$ and $\W(\e_1)$ is the skew-symmetric tensor associated with $\e_1$.
It follows from \eqref{eq:y_conical} and \eqref{eq:rotation} that  for all $\vartheta\in[0,\alpha]$ the polar angle of the first immersed sector is given by 
\eqref{eq:psi_star} in the main text.

The appropriate value of $\beta$ is determined by requiring that the plane spanned by $\e(\alpha)$ and $\E(\alpha)$ is the vertical plane used to reflect the first immersed sector in step~\ref{step:4} of our construction (see Fig.~\ref{fig:step07}). Formally, this geometric requirement demands that  $\proj(\e_3)\E(\alpha) = \delta\e(\alpha)$,  for some $\delta\in[0,1]$, where $\proj(\e_{3})$ is the projection onto the plane orthogonal to $\e_{3}$, that is,
	\begin{equation}\label{eq:beta_request}
\cos(\mu\alpha)\vc{e}_{1}+\sin(\mu\alpha)\cos\beta\vc{e}_{2}=
	\begin{cases}
	\delta(\cos\alpha\e_{1}+\sin\alpha\e_2) &\text{if }\mu\alpha<\frac{\pi}{2},\\
	\zero &\text{if }\mu\alpha=\frac{\pi}{2}.
	\end{cases}
	\end{equation}
Solving \eqref{eq:beta_request}, we easily find that $\beta$ is given by \eqref{eq:beta}, while
\begin{equation}
\label{eq:delta}
\delta=\frac{\cos\mu\alpha}{\cos\alpha}.
\end{equation}
Since, by \eqref{eq:y_conical}, $\E(\vt)\cdot\e_1=\sin\psi\cos(\vp+\vt)$ (see Fig.~\ref{fig:step06}), from \eqref{eq:rotation} we also retrieve \eqref{eq:phi_star}.

Finally, we note that the function $\psi^\ast$ can be extended continuously to the entire $\real$ by making it periodic with period $4\alpha$ and oscillating symmetrically about $\frac\pi2$. The extended function $\psi$ is represented as follows in one period,
\begin{equation}\label{eq:psi_extended}
\psi(\vartheta) = \left\{
\begin{aligned}
&\psi^\ast(\vartheta) \qquad &&\text{if }\vartheta\in[0,\alpha],\\
&\psi^\ast(2\alpha-\vartheta) \qquad &&\text{if }\vartheta\in[\alpha,2\alpha],\\
&\pi-\psi^\ast(\vartheta-2\alpha) \qquad &&\text{if }\vartheta\in[2\alpha,3\alpha],\\
&\pi-\psi^\ast(4\alpha-\vartheta) \qquad &&\text{if }\vartheta\in[3\alpha,4\alpha].
\end{aligned}
\right.
\end{equation}
Similarly, the periodic $C^1$-extension of $\vp^\ast$ compatible with \eqref{eq:phi_prime} is given  by
\begin{equation}\label{eq:phi_extended}
\varphi(\vartheta) = \left\{
\begin{aligned}
&\varphi^\ast(\vartheta) \qquad &&\text{if }\vartheta\in[0,\alpha],\\
&-\varphi^\ast(2\alpha-\vartheta), \qquad &&\text{if }\vartheta\in[\alpha,2\alpha],\\
&\varphi^\ast(\vartheta-2\alpha) \qquad &&\text{if }\vartheta\in[2\alpha,3\alpha],\\
&-\varphi^\ast(4\alpha-\vartheta), \qquad &&\text{if }\vartheta\in[3\alpha,4\alpha].
\end{aligned}
\right.
\end{equation} 
Since both $\sin\psi$ and $\psi'^2$ are $2\alpha$-periodic functions, by \eqref{eq:phi_prime} so is also $\vp$. Moreover, 
it is easy to check that 
\begin{equation}
\label{eq:periodic_values}
\psi'^2((2k+1)\alpha)=\frac{\mu^2}{\sin^2\alpha}(\cos^2\alpha-\cos^2(\mu\alpha))\quad\text{and}\quad\vp(k\alpha)=0\quad\text{for all}\ k\in\mathbb{N},
\end{equation}
as illustrated in Fig.~\ref{fig:psi_phi_shapes}.

%

\end{document}